\documentclass[aps,prc,twocolumn,superscriptaddress,showpacs,nofootinbib]{revtex4-2}
\usepackage{graphicx}
\usepackage{amsmath,amssymb}
\usepackage{bm}
\usepackage{xcolor}
\usepackage[colorlinks,citecolor=blue]{hyperref}
\usepackage{orcidlink}

\newcommand{\Qcr}{Q_\mathrm{b}}
\newcommand{\Qgs}{Q_\mathrm{gs}}
\newcommand{\Qsp}{Q_\mathrm{sp}}
\newcommand{\Egs}{E_\mathrm{gs}}
\newcommand{\Vpot}[1]{V_{#1}}

\begin{document}

\title{Isoenergetic description of induced fission pathways within energy-density functional theory}

\author{Alan A. Dzhioev\orcidlink{0000-0003-0642-1079}}
\email{dzhioev@theor.jinr.ru}
\affiliation{Bogoliubov Laboratory of Theoretical Physics, JINR, 141980, Dubna, Russia}

\author{N. V. Antonenko\orcidlink{0000-0002-3666-3389}}
\affiliation{Bogoliubov Laboratory of Theoretical Physics, JINR, 141980, Dubna, Russia}
\affiliation{Tomsk Polytechnic University, 634050, Tomsk,  Russia}

\date{\today}

\begin{abstract}

A thermodynamically consistent description of induced fission pathways in the superheavy nucleus $^{296}$Lv is presented within the framework of nuclear energy-density functional theory. Using self-consistent finite-temperature Hartree-Fock-Bogoliubov calculations with the Skyrme-type energy-density functional SkM$^*$, we derive and compare effective potentials corresponding to different thermodynamic processes---isothermal ($T=\text{const}$), isentropic ($S=\text{const}$), and isoenergetic ($E=\text{const}$)---as functions of quadrupole deformation and excitation energy. At  the same amount of excitation energy the isoenergetic description predicts the highest fission barrier  and the largest damping factor.  The suppression of the isoenergetic fission barrier is analyzed by examining how the driving force changes with increasing excitation energy and by studying the contribution of the nonpotential term to the effective potential. Within the framework of the three thermodynamic schemes considered, the different behavior of temperature and entropy along the fission pathway is also emphasized.
A transition of the effective level-density parameter from a deformation-sensitive quantity at low excitation energies to a nearly constant value at high energies is observed, in line with the Fermi-gas model expectations.

\end{abstract}


\maketitle

\section{Introduction}

Understanding the fission dynamics of heavy and superheavy nuclei is a key challenge in modern nuclear theory, with important implications for both fundamental nuclear structure studies and applications in nuclear astrophysics and reactor physics~\cite{PRC79_Moller, RepProgPhys81_Schunck,RepProgPhys81_Andreyev,RepProgPhys81_Schmidt,ADNDT138_Jachimowicz,Schunck2023}.
A particularly important aspect is the accurate modeling of fission barriers at finite excitation energy, which govern the stability and survival probability of compound nuclei formed in heavy-ion fusion reactions~\cite{PRC65_Itkis,PRC76_Loveland,EPJA60_Rahmatinejad}.
Reliable theoretical predictions of these barriers require detailed knowledge of the potential energy surface with probable  fission pathways. The evolution of the effective potential landscape with excitation energy offers valuable insight into the dynamics of the fission process, revealing how shell effects, pairing correlations, and collective motion influence the formation and suppression of the fission barrier in excited compound nuclei. Indeed, with increasing excitation energy of the fissionable nucleus the Fermi surface is blurred.
As a result, the microscopic effects (shell and pairing corrections) change altering the topology of the potential energy surface and modifying the fission pathways.  In Ref.~\cite{EPJA60_Rahmatinejad}, the change
of competition between the fission pathways was revealed with increasing excitation energy.

The fission process can be related to one of three thermodynamic processes: isothermal ($T=\text{const}$), isentropic ($S=\text{const}$), or isoenergetic ($E=\text{const}$).
The isothermal process is commonly employed in the analysis of compound nucleus fission~\cite{ NPA264_Sauer,JPhG13_Okolowicz,IJMPE18_Martin, PRC91_Schunck, JPhConfSer436_Schunck}. It implies that the nucleus evolves along the fission path while maintaining a constant temperature $T$.
The isothermal process is computationally the most straightforward to handle. The finite-temperature mean-field equations are derived by minimizing the grand potential at a fixed  $T$~\cite{Goodman_NPA352} (see also Ref.~\cite{Schunck_book} for a detailed discussion).
It is important to note, however, that the assumption of constant temperature inherently implies the presence of an external thermal reservoir---a condition that clearly does not hold for an isolated nucleus. In contrast, this assumption may be more reasonable in the context of stellar matter, where nuclei are embedded in a nucleon vapor and interact with a thermal bath of photons~\cite{PPN53_1_Dzhioev,PPN53_2_Dzhioev,PPN53_3_Dzhioev}.
In Refs.~\cite{NPA335_Duebel,PRL102_Pei, PRC80_Sheikh, NPA834_Pei,PRC87_McDonnell}, it was pointed out that, in the absence of heat exchange with the thermal reservoir, the motion along the collective coordinate $Q$ can be considered adiabatic. Therefore, an isentropic process--that is, one occurring at constant entropy $S$---provides a more realistic description than the isothermal assumption.

As argued in~\cite{RepProgPhys81_Schmidt},  treating fission as an isentropic process is not a generally valid assumption since the laws of thermodynamics have a statistical nature and thermodynamic quantities, such as the entropy, are subject to fluctuations that become sizable in microscopic systems like nuclei. It was also emphasized that the evolution of the entropy plays a decisive role in the fission process and the concentration of a large amount of energy into the collective degrees of freedom at the barrier leads to a decrease of the thermal energy and induces a reduction of the entropy.

For a nucleus, the observable quantity is the energy, not the temperature or the entropy. Since the nucleus is an isolated system, its total energy $E$ remains constant until it deexcites via the emission of nucleons or photons. As the deformation changes, this constant total internal energy is redistributed between the static deformation  energy, and the  thermal energy.
For these reasons, an isoenergetic process  appears to be a more physically relevant approach for describing the evolution along the collective coordinate. The isoenergetic description of the fission process is widely used within the macroscopic-microscopic framework~\cite{PRC88_Randrup, EPJA60_Rahmatinejad} and in stochastic approaches to fission dynamics based on Langevin equations~\cite{NPA556_Frobrich, PhysRep292_Frobrich}.

Among various theoretical approaches, energy-density functional theory (EDFT) has emerged as a powerful and predictive framework for describing ground-state properties, collective motion, and thermal effects across the nuclear chart~\cite{Schunck_book,APhPolB44_Schunck}. In this work, we employ EDFT to develop a thermodynamically consistent description of the fission pathway in excited compound nuclei, focusing on the superheavy nucleus $^{296}$Lv as an example. While most existing EDFT studies of induced fission  rely on either isothermal~\cite{IJMPE18_Martin, PRC91_Schunck, JPhConfSer436_Schunck} or isentropic~\cite{PRL102_Pei, PRC80_Sheikh, NPA834_Pei,PRC87_McDonnell} descriptions, we argue that the isoenergetic process  provides a more physically relevant framework for modeling fission pathways of isolated excited nuclei. To the best of our knowledge, the isoenergetic condition has not yet been employed to derive the effective potential within self-consistent EDFT calculations.

This paper is structured as follows: In Sec. II, we outline the formalism and thermodynamic framework used in our calculations. Section III presents the calculated results for $^{296}$Lv, including the evolution of the fission barrier, driving force, temperature, entropy, and level-density parameters with deformation and excitation energy. Section IV contains concluding remarks and perspectives for future work.

\section{Formalism}

We consider excited compound nuclei formed in complete fusion reactions as isolated quantum systems, for which no coupling to external fields or to an external thermal reservoir exists. As usual, we assume that during the fission process the nucleus slowly evolves by changing its quadrupole deformation $Q$, and at each deformation the local equilibrium is achieved, which is characterized by the total internal energy $E$. In the statistical description of the fission process, the deformation and the energy are given by ensemble-average values of the nuclear Hamiltonian, $E = \langle \hat{H} \rangle$, and the respective quadrupole moment operator, $Q = \langle \hat{Q} \rangle$. Within the grand canonical description, the particle number is also given in the form of an average
$N_\tau = \langle \hat{N}_\tau \rangle$ ($\tau = n,\,p$).

To describe local equilibrium properties of a compound nucleus, we are looking for the normalized density operator $\hat D$ ($\mathrm{Tr}\hat D=1$) which satisfies the constraints
\begin{eqnarray}\label{D_constraints}
  &E =\mathrm{Tr}(\hat D\hat H),\notag\\
  &Q = \mathrm{Tr}(\hat D\hat Q),\notag\\
  &N_\tau = \mathrm{Tr}(\hat D\hat N_\tau).
\end{eqnarray}
This problem can be solved with the help of the "maximum statistical entropy principle" (see, for example, Chapters 3 and 4 in Ref.~\cite{Balian_book}). Namely, the maximization of the entropy $S=-\mathrm{Tr}(\hat D\ln\hat D)$ under constraints~\eqref{D_constraints} leads to the following expression for the density operator:
\begin{equation}\label{density_D}
 \hat D = \frac{1}{Z} e^{-\beta\hat H+\alpha_n\hat N_n + \alpha_p\hat N_p + \lambda  \hat Q},
\end{equation}
where $\beta$, $\alpha_\tau$, and $\lambda$ are Lagrangian multipliers, and conditions~\eqref{D_constraints} determine them as functions of the averages $E$, $N_\tau$, and $Q$.

The partition function
\begin{equation}\label{Z_function}
  Z = \mathrm{Tr}\left( e^{-\beta\hat H + \alpha_n\hat N_n + \alpha_p\hat N_p + \lambda \hat Q} \right)
\end{equation}
contains all thermodynamic information about the system and satisfies the following fundamental relations:
\begin{eqnarray}\label{partial_Z}
  \frac{\partial}{\partial\beta} \ln Z &=& -E, \notag \\
  \frac{\partial}{\partial\alpha_\tau} \ln Z &=& N_\tau, \notag \\
  \frac{\partial}{\partial\lambda} \ln Z &=& Q.
\end{eqnarray}
Using the definition of entropy, one derives the relation
\begin{equation}
  S = \ln Z + \beta E - \alpha_n N_n - \alpha_p N_p - \lambda Q.
\end{equation}
By taking differentials and applying Eqs.~\eqref{partial_Z}, we obtain
\begin{equation}\label{dS1}
    dS = \beta\,dE - \alpha_n\,dN_n - \alpha_p\,dN_p - \lambda\,dQ.
\end{equation}
This equation expresses how the local equilibrium entropy changes in response to infinitesimal variations in the parameters $E$, $N_\tau$, and $Q$, which fully characterize the local equilibrium state of the nucleus.

In the following, we assume that the average number of particles in the nucleus remains constant during the deformation process. Then, by introducing
$T = \beta^{-1}$ and $\kappa= \lambda / \beta$, we  can recast Eq.~\eqref{dS1} into a form analogous to the first law of thermodynamics:
\begin{equation}\label{dE}
    dE = T\,dS + \kappa\,dQ.
\end{equation}
Here, $T$ and $\kappa$ acquire the physical interpretations of temperature and collective driving force, respectively---quantities that determine the conditions necessary to maintain nuclear equilibrium at given values of total energy and deformation. However, since the nucleus is an isolated quantum system, Eq.~\eqref{dE} should not be interpreted in terms of heat exchange or mechanical work. Rather,  this equation expresses how the internal energy change $dE$ is distributed between the thermal component $T\,dS$ and the energy $\kappa\,dQ$ associated with the evolution of the collective coordinate.

Using the definition of the Helmholtz free energy, $F = E - TS$, the driving force $\kappa$ is expressed in terms of partial derivatives of three distinct thermodynamic quantities:
\begin{eqnarray}\label{kappa_def}
 \kappa(Q,T) &=& \frac{\partial F(Q,T)}{\partial Q}, \notag \\
 \kappa(Q,S) &=& \frac{\partial E(Q,S)}{\partial Q}, \notag \\
 \kappa(Q,E) &=& -T(Q,E)\,\frac{\partial S(Q,E)}{\partial Q}.
\end{eqnarray}
These expressions are equivalent under the condition that the variables $T$, $S$, and $E$ are functionally related. Specifically, if we consider $T=T(Q,E)$, $S=S(Q,E)$, or equivalently, any of the alternative dependencies such as $E=E(Q,T)$, $S=S(Q,T)$, or $E=E(Q,S)$, $T=T(Q,S)$, then
\begin{equation}\label{kappa_eq}
 \kappa(Q,T) = \kappa(Q,S) = \kappa(Q,E).
\end{equation}
This equivalence ensures that the driving force is well defined irrespective of the chosen set of independent variables.

To define the effective potential $\Vpot{X}(Q)$, for $X = T,\,S,\,E,$\footnote{For clarity, here and below the subscript indicates the thermodynamic process to which the effective potential refers.} such that $\Vpot{X}(Q) = \kappa(Q,X)\,dQ$ represents the infinitesimal change in the energy associated with collective motion between two neighboring points $Q$ and $Q + dQ$, it is necessary to specify the thermodynamic process governing the evolution of the nucleus along the fission pathway.

Under isothermal conditions ($T=\text{const}$)  the total derivative of the free energy with respect to the collective coordinate $Q$ can be used directly in Eq.~\eqref{kappa_def}. Consequently, up to an additive deformation-independent constant, the isothermal effective potential coincides with the Helmholtz free energy, i.e. $\Vpot{T}(Q) = F(Q, T)$.
In this thermodynamic framework, both the internal energy and entropy become the functions of deformation and temperature, $E = E(Q, T)$ and $S = S(Q, T)$. For a cold nucleus ($T = 0$), the isothermal effective potential $\Vpot{T}(Q)$ reduces to the deformation energy $V(Q)$, which represents the minimum internal energy achievable for a given $Q$.

In the isentropic consideration ($S=\text{const}$), the temperature varies with deformation, and from Eq.~\eqref{kappa_def} it follows that the internal energy defines the deformation-dependent part of the effective potential:
$$
\Vpot{S}(Q) = E(Q, S).
$$
As in the isothermal case, for a cold nucleus ($S = 0$), the isentropic effective potential $\Vpot{S}(Q)$ reduces to the deformation energy $V(Q)$.

\begin{figure}[htbp]
\centering
\includegraphics[width=0.85\linewidth]{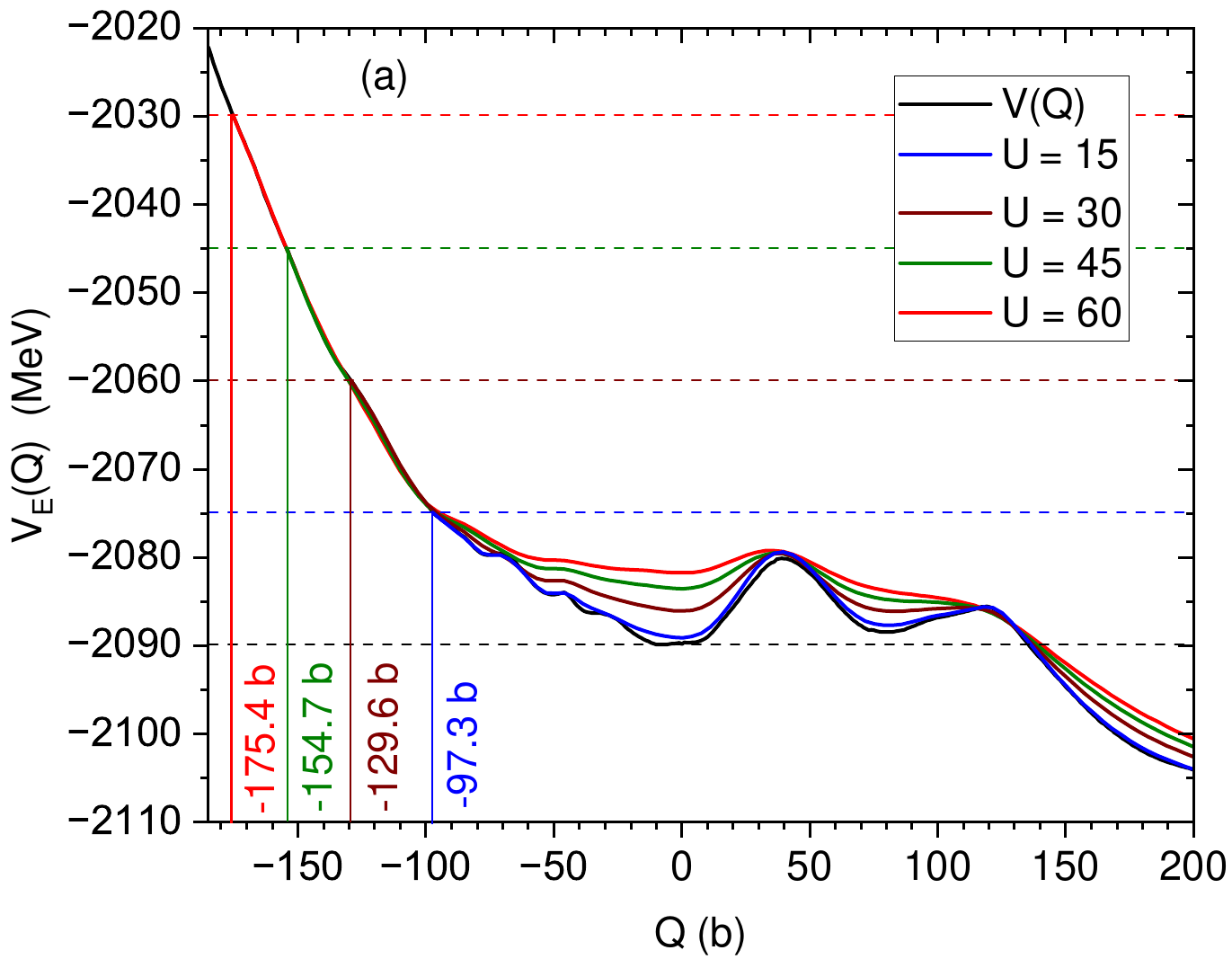}

\includegraphics[width=0.85\linewidth]{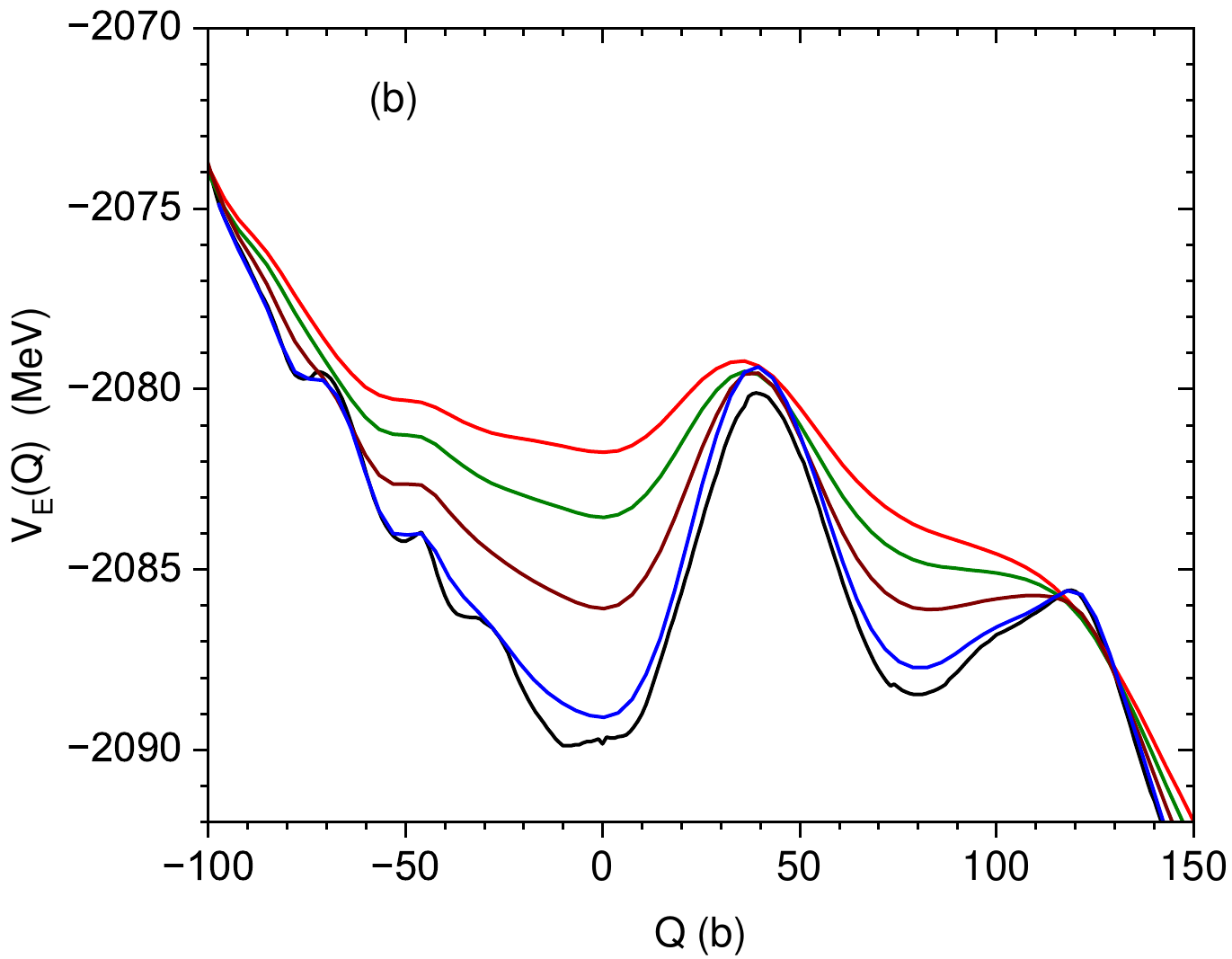}
\caption{(a) The deformation energy $V(Q)$ is compared with the isoenergetic effective potentials $\Vpot{E}(Q)$ at selected excitation energies $U$ (in MeV). The dashed lines indicate the total internal energy $E = \Egs + U$, where $\Egs = -2089.88$\,MeV is the Hartree-Fock-Bogoliubov ground-state binding energy obtained using the SkM* energy density functional. It is worth recalling that for a fixed excitation energy $U$, the boundary $\Qcr$ of the isoenergetic potential is determined by the condition $V(\Qcr) = E$. The vertical lines mark the values of $\Qcr$ corresponding to the selected excitation energies $U$. (b) A close-up view of the upper panel, focusing on the fission barrier region.}
    \label{potential_E}
\end{figure}

Let $U$ denote the excitation energy, so that the total internal energy of the excited nucleus is given by $E = \Egs + U$, where $\Egs$ is the ground-state energy defined as the global minimum of the deformation energy $V(Q)$.
In an isoenergetic process, the deformation energy cannot exceed the total energy. Therefore, the condition $V_E(Q) \leq E$ determines the physically allowed domain of the deformation coordinate $Q$. This is in contrast to isothermal and isentropic descriptions, in which the internal energy is not conserved, and, in principle, $Q$ can take arbitrary values.
Assuming that the nuclear excitation energy $U$ exceeds the height of the fission barrier, there exists a unique domain boundary $\Qcr$ characterized by $V(\Qcr)=E$ (see Fig.~\ref{potential_E}). Typically,  $\Qcr$ corresponds to an oblate nuclear shape. Smaller $Q$ are energetically forbidden, while at $Q > \Qcr$ the local excitation energy $U(Q) = E - V(Q)$ is positive. At $\Qcr$ the entire excitation energy is completely converted to deformation energy and it is natural to impose the boundary condition $\Vpot{E}(\Qcr) = E$.
Then, integrating $d\Vpot{E}= - T\dfrac{dS}{dq}dq$
from $\Qcr$ to $Q$, we obtain the expression for the isoenergetic effective potential:
\begin{equation}\label{V_E}
  \Vpot{E}(Q) = E - \int^Q_{\Qcr} T(q,E)\frac{d S(q,E)}{d q} dq.
\end{equation}
Note that $\Qcr$ depends on the excitation energy $U$ and at $U = 0$, the effective potential is defined only at the ground-state deformation $\Qgs$, where it reduces to $\Vpot{E}(\Qgs) = \Egs$.

It can be argued that since the integrand in~\eqref{V_E} is not a full differential, the isoenergetic effective potential is not a function of the state alone but also depends on the fission pathway. For example, $\Vpot{E}(Q)$  can differ for axial symmetric and triaxial pathways, even if the initial and final states are identical \cite{EPJA60_Rahmatinejad}. However, for fully symmetry-unrestricted calculations, the fission pathway is uniquely determined by the condition of maximum entropy at a fixed quadrupole moment. In such cases, the isoenergetic effective potential becomes a function of state.

It is instructive to compare the expressions for the effective potentials $\Vpot{X}(Q)$ ($X=T,\,S,\,E$) within the Fermi gas model. In this model, the local entropy and temperature are given by $S(Q) = 2\sqrt{a(Q)\,U(Q)}$ and $T(Q) = \sqrt{U(Q)/a(Q)}$, where $a(Q)$ is the deformation-dependent level-density parameter and $U(Q) = E - V(Q)$ is the local excitation energy.
Using the general definitions of the effective potentials introduced above, we obtain:
\begin{eqnarray}
  \Vpot{T}(Q)  &=& V(Q) - a(Q)T^2,\notag \\
  \Vpot{S}(Q)  &=& V(Q) + a(Q)T^2(Q), \quad \big(T(Q) = S/2a(Q)\big),\notag \\
  \Vpot{E}(Q) &=& V(Q) - \int_{\Qcr}^Q T^2(q)\frac{da(q)}{dq} dq.
\end{eqnarray}
Notably, if the level density parameter $a(Q)$ is independent of deformation, the effective potentials differ from the deformation energy $V(Q)$ only by a constant shift. As a result, the three thermodynamic descriptions---isothermal, isentropic, and isoenergetic---become physically equivalent in this case.

In the present work, we study the effective potentials as functions of excitation energy within the mean-field approximation by solving the finite-temperature Hartree-Fock-Bogoliubov (FTHFB) equations. These equations are derived using EDFT
with the Skyrme-type particle-hole interaction. The proof of thermodynamic consistency of the FTHFB approximation is provided in Refs.~\cite{PRC93_Alhassid,EPJA57_Ryssens}, where the validity of relations~\eqref{partial_Z} is verified. This is a nontrivial observation, since the
mean-field  Hamiltonian depends self-consistently on the  Lagrangian multipliers. The validity of these relations within the FTHFB framework ensures that the grand-canonical entropy calculated in this approach satisfies fundamental thermodynamic identity~\eqref{dS1}.

The details of the derivation of the FTHFB equations are given in Refs.~\cite{JPhG19_Egido,Schunck_book}. Here we just  recall that in the FTHFB method, density matrix~\eqref{density_D} is approximated by a trial density matrix of the form
\begin{equation}
  \hat D_\mathrm{HFB} = \frac{e^{-\beta \hat K}}{Z_\mathrm{HFB}},~~Z_\mathrm{HFB}=\mathrm{Tr}\big(e^{-\beta \hat K}\big),
\end{equation}
where $\hat K=\hat K^\dag$ is a general quadratic form  in nucleon creation and annihilation operators (see Eq. (7.15) in Ref.~\cite{Schunck_book} or Eq.~(2.9) in Ref.~\cite{JPhG19_Egido}).    The matrix elements of $\hat K$ play the role of variational parameters.
Then, the  equilibrium state of a hot nucleus at constant temperature $T$ and chemical potential $\mu$ is obtained from the minimization of
the grand canonical potential $\Omega = E-TS-\mu N$ with respect to variations $\delta\hat K$~\cite{JPhG19_Egido,Schunck_book}.   The resulting FTHFB equations take the same form as the Hartree-Fock-Bogoliubov (HFB) equations at $T=0$, only the one-body particle and pairing density matrices now depend on the Fermi-Dirac occupation  of quasiparticle states. To constrain the quadrupole  moment $Q$ the quantity to be minimized is
$\Omega'= \Omega - \kappa \langle\hat Q\rangle$. By performing zero-temperature constrained HFB calculations, we obtain the deformation energy curve $V(Q)$ within the mean-field approximation. The global minimum of $V(Q$) yields the HFB binding energy $\Egs$ and the corresponding ground-state deformation $\Qgs$.

Once the FTHFB density operator is determined, it can be used to evaluate the physical quantities, like  the internal energy $E(Q,T)$ or the entropy $S(Q,T)$, as a function of temperature and quadrupole moment. The Helmholtz free energy $F(Q,T)=E(Q,T)-TS(Q,T)$ at fixed $T$ yields the effective potential in the isothermal description of the fission process. Passage to  isentropic and isoenergetic  descriptions is performed by constraining the FTHFB calculations for a number of temperatures and then using $E(Q,T)$ and $S(Q,T)$ to reconstruct $E(Q,S)$, $T(Q,S)$ and $S(Q,E)$, $T(Q,E)$ numerically by using interpolation.

In the present study, the calculations are carried out for $^{296}$Lv using the axial-symmetric EDFT solver {HFBTHO}~\cite{CPC184_Stoitsov}, which (i) solves the FTHFB equations in the one-center harmonic oscillator (HO) basis and (ii) allows to perform multiconstraint calculations using the linear constraint method.
The results presented below are obtained with the Skyrme  functional SkM* that has been optimized at large deformations~\cite{NPA386_Bartel}. The details of the present HFBTHO calculations follow Refs.~\cite{APhPolB44_Schunck,PRC90_Schunck,PRC91_Schunck}. Namely, in the pairing channel we employ the density-dependent  contact interaction with mixed volume-surface character~\cite{EPJA15_Dobaczewski}. The parameters of the pairing interaction, i.e., the energy cutoff and the pairing strengths for  protons and neutrons,  were fitted to reproduce the three-point odd-even mass difference in $^{240}$Pu.
The calculations were performed using the expansion of HFB wave functions over the lowest 1100 deformed oscillator states from up to 31 oscillator shells.
The deformation and the oscillator frequency of the HO basis were parametrized as a function of quadrupole moment according to the method in Refs.~\cite{APhPolB44_Schunck,PRC90_Schunck}.

In the HFBTHO solver, the Lagrangian multiplier $\kappa$ conjugate to the quadrupole moment $Q$ is represented by the variable \texttt{multLag(2)}. By performing constrained FTHFB calculations over a grid of temperatures and deformations, we obtain a two-dimensional surface $\kappa(Q, T)$. Using the corresponding energy $E(Q, T)$ and entropy $S(Q, T)$, this surface is numerically transformed into other thermodynamic variables, such as $\kappa(Q, S)$ or $\kappa(Q, E)$.
The resulting driving force $\kappa(Q, X)$ ($X=T,\,S,\,E$) is then used to compute the effective potential $\Vpot{X}(Q)$ via numerical integration. This approach is particularly advantageous for isoenergetic calculations, as it avoids the need for numerical differentiation of the entropy.
In this way, the isoenergetic effective potential is expressed as
\begin{equation}\label{V_E2}
  \Vpot{E}(Q) = E + \int^{Q}_{\Qcr} \kappa(q, E)\, dq.
\end{equation}
The condition $\kappa(Q, E) = 0$ defines the stationary points of both the effective potential and the entropy, which are of particular interest in the analysis of fission pathways.

\section{Calculated results and discussions}

\subsection{Effective potential along fission pathway}

As previously stated, the present study employs axially symmetric numerical calculations for $^{296}$Lv using the SkM$^*$ parametrization of the Skyrme EDFT. Reflection-asymmetric degrees of freedom, such as octupole deformations, are omitted from the analysis, as they become significant only beyond the first fission barrier -- the region of primary interest in this work~\cite{PRC85_Abusara, PRC92_Scamps, PRC93_Zhao}. Constrained FTHFB calculations were carried out over a range of quadrupole deformation values $Q$ from $-200$ to $+200$~b with a step size of 1~b. At each deformation value $Q$, the temperature $T$ was incremented in steps of 0.05~MeV starting from zero, until the total internal energy $E(Q,T)$ exceeded the ground-state energy $\Egs$ by 70 MeV. The transformation from the $(Q,T)$ grid to the $(Q,E)$ and $(Q,S)$ representations was performed using bilinear interpolation. Cubic spline interpolation was applied to smooth the curves shown in all the figures.

Figure~\ref{potential_E} compares the deformation energy curve $V(Q)$ for $^{296}$Lv, calculated using the SkM* parametrization, with the effective potentials $\Vpot{E}(Q)$ at selected excitation energies: $U = 15,\,30,\,45$, and $60$\,MeV.
In the ground state of the cold nucleus $^{296}$Lv, the SkM* functional predicts a slightly oblate shape with the quadrupole moment $\Qgs \approx -7.7$\,b. The first saddle point is located at $\Qsp \approx 39$\,b, and the corresponding fission barrier height is $B_f \approx 9.8$\,MeV.
All these predictions are in excellent agreement with the microscopic EDFT calculations using the Gogny D1S interaction~\cite{PRC86_Warda,NPA944_Baran}, confirming the robustness of the theoretical approach.
In  $^{296}$Lv, both the SkM* and Gogny D1S  reflection-symmetric  calculations  result in the  second (outer) fission barrier at  $Q\approx 120$\,b, which separates the   fission isomer from the scission region.
As demonstrated in Refs.~\cite{PRC86_Warda,NPA944_Baran}, the inclusion of reflection-asymmetric degrees of freedom completely removes the second fission barrier in $^{296}$Lv, while leaving the deformation energy curve essentially unchanged at $Q< 70$\,b.

The $Q$ dependence of the isoenergetic effective potential shows that the nucleus  becomes spherical  ($\Qgs^* = 0$) even at low excitation energies, while the position $\Qsp$ of the saddle point remains largely independent of $U$. Our calculations reveal that the isoenergetic fission barrier,
which keeps the nucleus bound, undergoes dramatic suppression with excitation energy, decreasing from
$B_f\approx 9.7$\,MeV at $U=15$\,MeV to merely 2.5\,MeV at $U=60$\,MeV.  The outer barrier shows even greater sensitivity, being completely washed out at
$U=30$\,MeV. As evident from the figure, the reduction of barriers primarily occurs because the effective potential flattens near the minima with increasing excitation energy, while $\Vpot{E}(Q)$ remains largely unchanged at the saddle points. This finding validates the microscopic-macroscopic approach, which predicts that the fission barrier is mainly determined by the negative shell-correction energy at
ground-state deformation, which disappears with increasing excitation energy \cite{EPJA60_Rahmatinejad}.
In contrast, weak shell effects at the saddle point result in an effective potential that varies only slightly with excitation energy.

The dashed horizontal lines in the upper panel of Fig.~\ref{potential_E} denote the total energies $E$ for  each value  of $U$. The variation of the effective excitation energy $U_E(Q)=E-\Vpot{E}(Q)$  with nuclear deformation demonstrates the conversion between thermal energy and collective motion energy as the nucleus changes its shape at constant  $E$. Focusing on the fission barrier region, we find that the local excitation energy  $U(Q)=E-V(Q)$ is consistently higher than the thermal energy $U_E(Q)$ and the difference increases with excitation energy.

The interpretation of the effective excitation energy $U_E(Q)$ as a  thermal energy becomes more transparent when the entropy is formulated within the mean-field approximation:
\begin{equation}
  S = -\sum_i \big[ f_i \ln f_i + (1 - f_i) \ln (1 - f_i) \big],
\end{equation}
where $f_i$ are the quasiparticle occupation probabilities at the corresponding quasiparticle energies $\varepsilon_i$.
From this, it follows that
\begin{equation}
  T \frac{dS}{dq} = \sum_{i} \varepsilon_i \frac{df_i}{dq},
\end{equation}
and the effective excitation energy is rewritten as
\begin{equation}\label{Ueff}
  U_E(Q) = \sum_i f_i \varepsilon_i - \int^Q_{\Qcr} \sum_{i} f_i \frac{d\varepsilon_i}{dq} dq.
\end{equation}
Thus, within the mean-field approximation, the effective excitation energy corresponds to the energy $\Sigma_E(Q)$ of thermally excited quasiparticles, minus a correction term $R_E(Q)$, which accounts for the energy associated with the rearrangement of the quasiparticle spectrum during deformation.

\begin{figure}[htbp]
\centering
\includegraphics[width=0.85\linewidth]{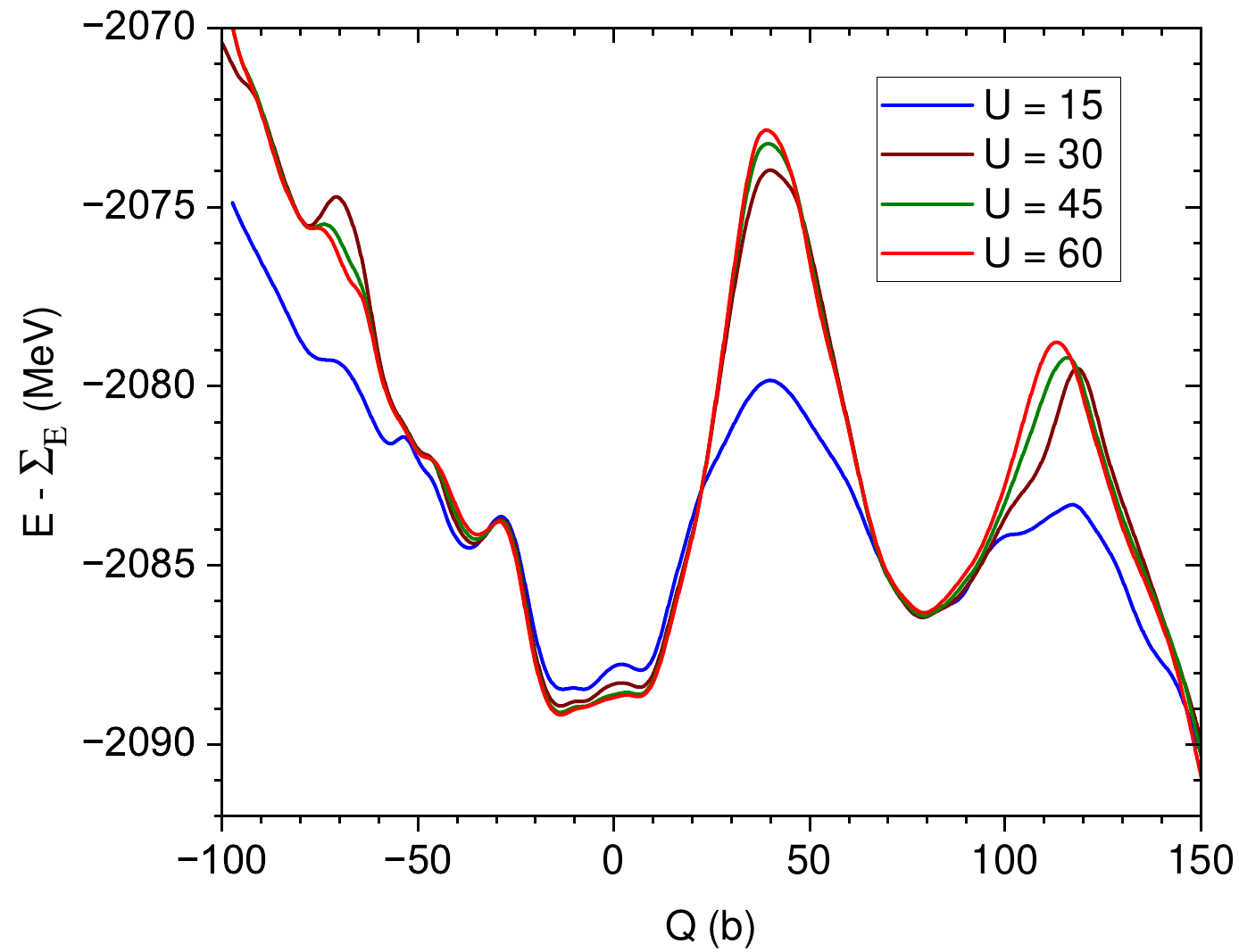}
    \caption{The potential component $E-\Sigma_E(Q)$ of the isoenergetic potential at indicated excitation energies $U$~(in MeV). }
    \label{Veff_pot}
\end{figure}

The above analysis demonstrates that, within the mean-field approximation, the nonpotential nature of $\Vpot{E}(Q)$ arises exclusively from the rearrangement term  $R_E(Q)$. To assess the contribution of this term  to the isoenergetic potential, we compare the effective potential $\Vpot{E}(Q)$  with its potential component $E-\Sigma_E(Q)$.  Figure~\ref{Veff_pot} displays $E-\Sigma_E(Q)$ at the same excitation energies as in Fig.~\ref{potential_E}, allowing for a direct comparison.  As seen in Fig.~\ref{Veff_pot}, the fission barrier persists even at high excitation energies when the rearrangement term $R_{E}(Q)$ is omitted. Moreover, $E-\Sigma_E(Q)$ retains the minimum at the ground-state oblate  deformation $\Qgs\approx -7.7$\,b.
These results suggest that the nonpotential term causes the fission barrier to decrease and drives the nucleus toward a spherical shape as the excitation energy increases. We also observe, that for sufficiently high excitation energies,  $E-\Sigma_E(Q)$ is almost independent of $U$.
This stabilization implies that an increase in excitation energy by $\Delta U$ results in an almost identical increase in the energy $\Sigma_E(Q)$ of quasiparticle excitations. A comparison of Figs.~\ref{potential_E} and~\ref{Veff_pot} reveals that in the minimum of isoenergetic potential, $\Qgs^*$, the contribution of the rearrangement term is positive and increases with excitation energy, while at the saddle point it is negative and becomes nearly independent of excitation energy at high~$U$.

\begin{figure}[htbp]
\centering
\includegraphics[height=0.7\linewidth]{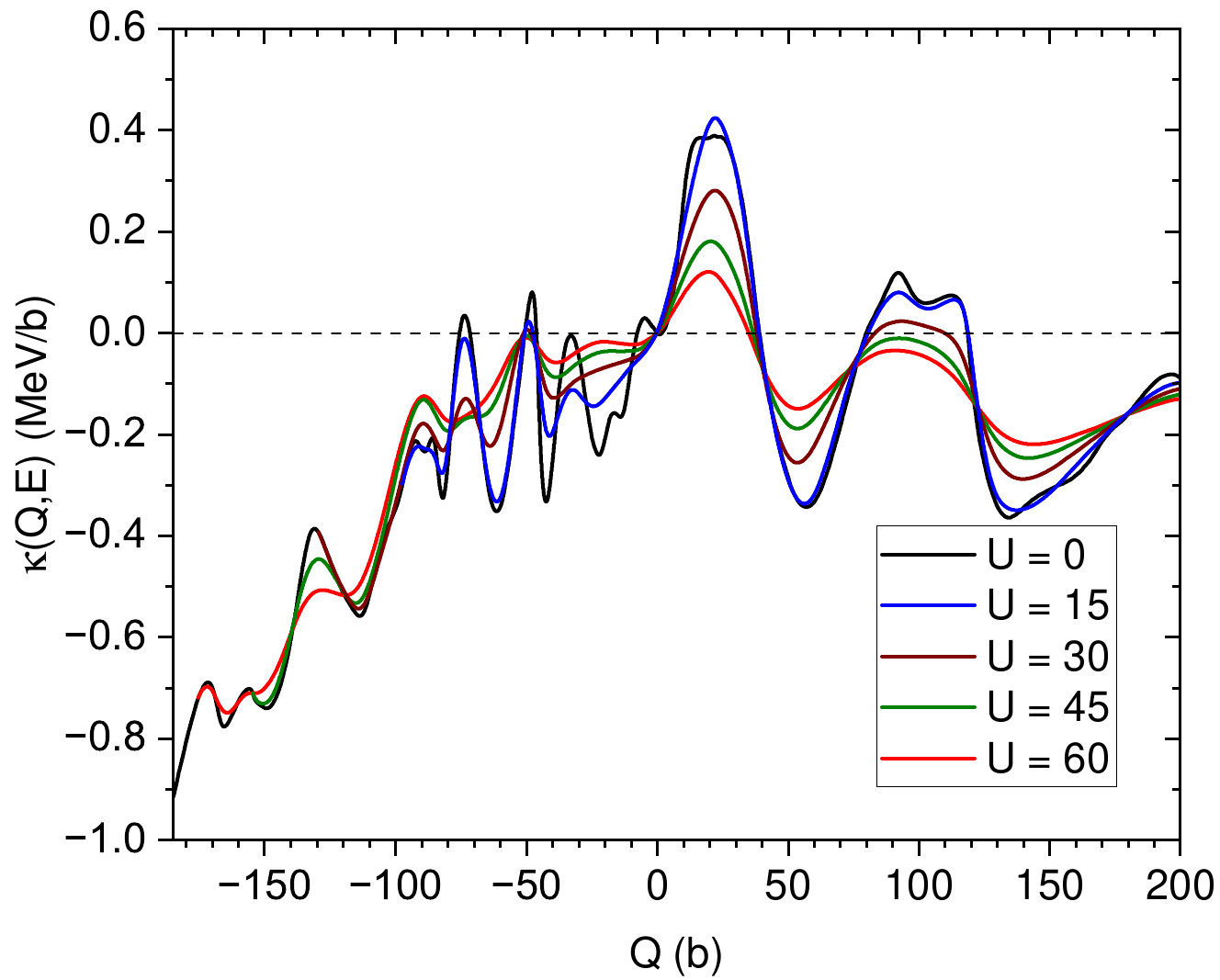}
\caption{The driving force $\kappa(Q,E)$ along the fission pathway of $^{296}$Lv at indicated excitation energies $U$ (in MeV).}
    \label{kappa_E}
\end{figure}

Let us now examine how the properties of the effective potential are related to the dependence of the driving force on excitation energy and deformation.   Stationary points of  $\Vpot{E}(Q)$  correspond to zeros of the driving force $\kappa(Q,E)$. The sign of the $\kappa(Q,E)$ derivative at these  points distinguishes between minima and saddle points of the effective potential. In Fig.~\ref{kappa_E}, we present $\kappa(Q,E)$ corresponding to the same excitation energies $U$ as those employed in Fig.~\ref{potential_E}. We also show the driving force $\kappa(Q)$  for  the cold nucleus $^{286}$Lv ($U=0$).   As seen, the driving force exhibits a complex dependence on the quadrupole deformation, with multiple extrema and stationary points  indicating variations in the effective potential landscape. At low excitation energies,  the driving force remains close to that for the cold nucleus. It reaches its most positive values between the stationary points $\Qgs^*$ and $\Qsp$, reflecting a strong restoring force that stabilizes the nucleus against fission.
The decrease in $\kappa(Q,E)$ between these stationary points with increasing excitation energy illustrates how
the fission barrier changes
\begin{equation}
 \frac{dB_f(E)}{dE} \approx \int^{\Qsp}_{\Qgs^*}\frac{\partial\kappa(Q,E)} {\partial E} dQ <0
\end{equation}
and how the fission process becomes increasingly probable as more energy is added to the system.

\begin{figure*}[htbp]
\centering
\includegraphics[width=0.75\columnwidth]{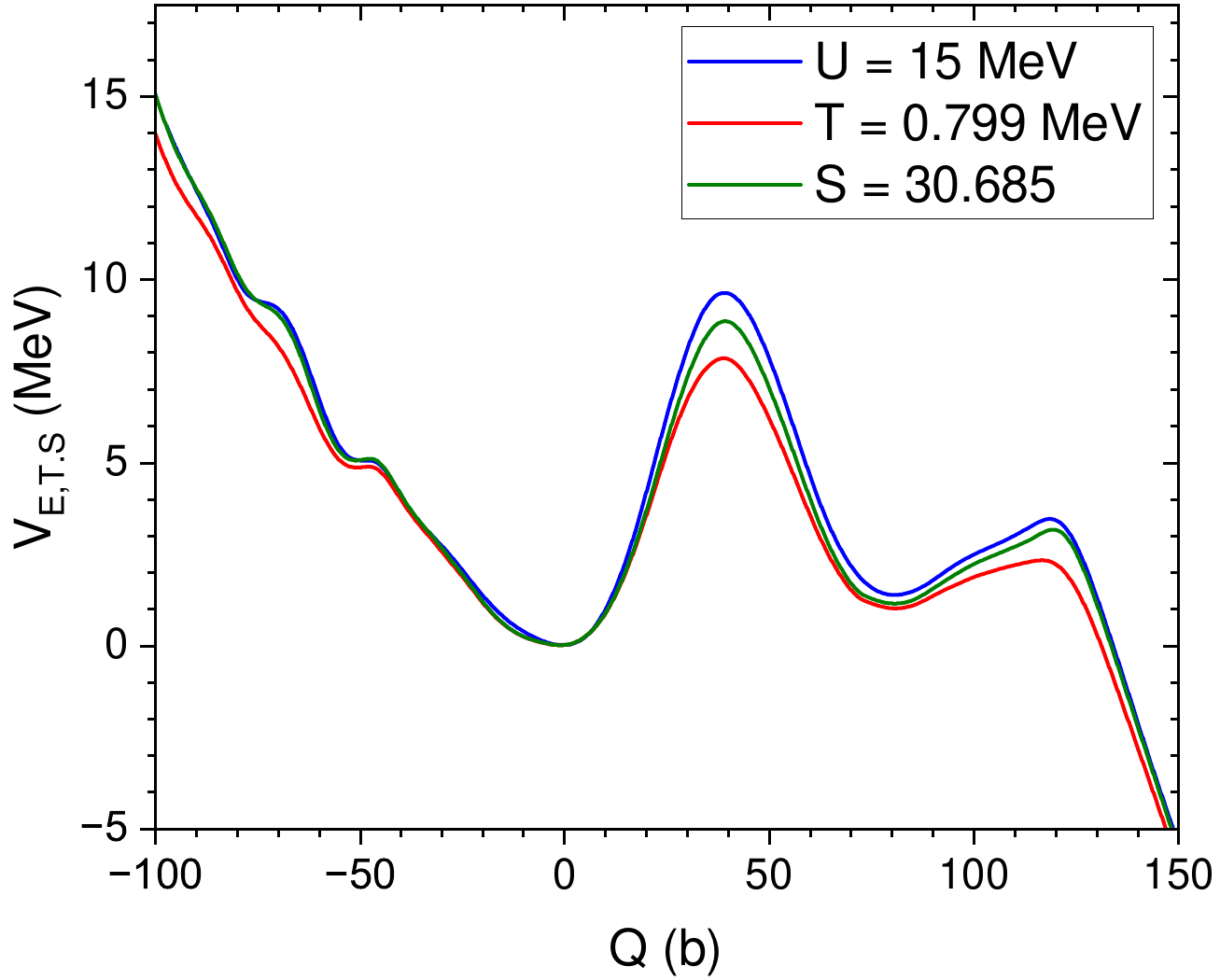}~~\includegraphics[width=0.75\columnwidth]{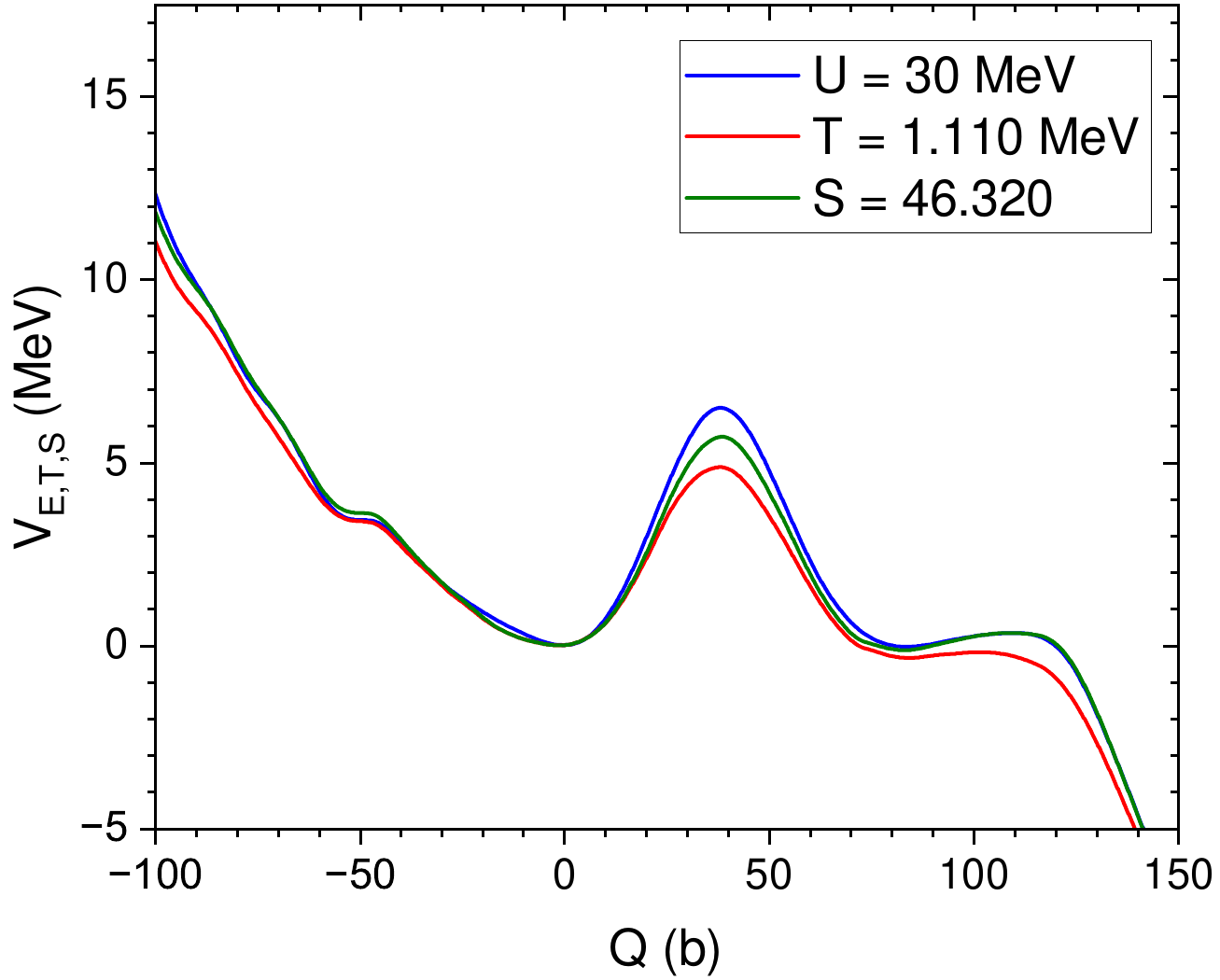}

\includegraphics[width=0.75\columnwidth]{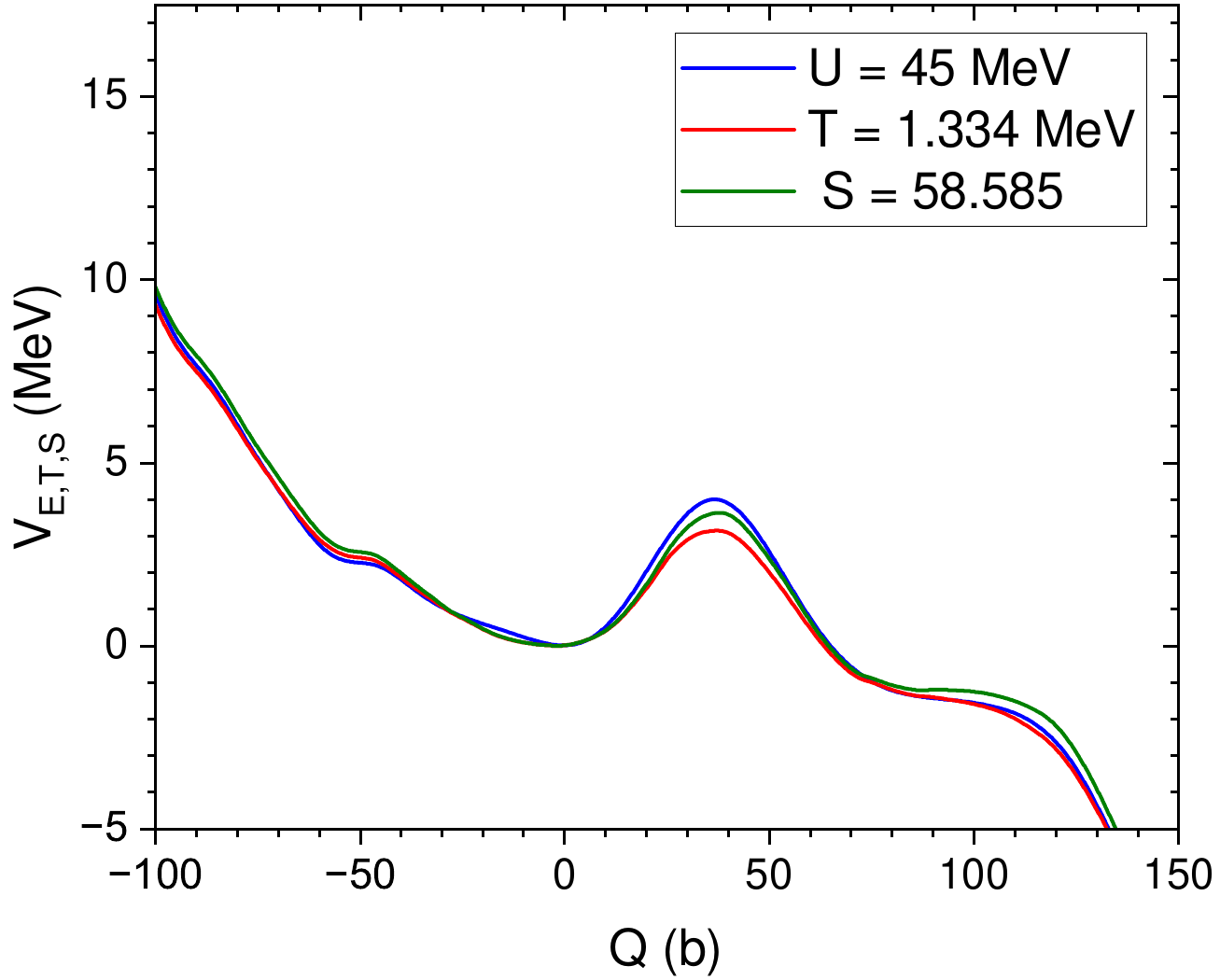}~~\includegraphics[width=0.75\columnwidth]{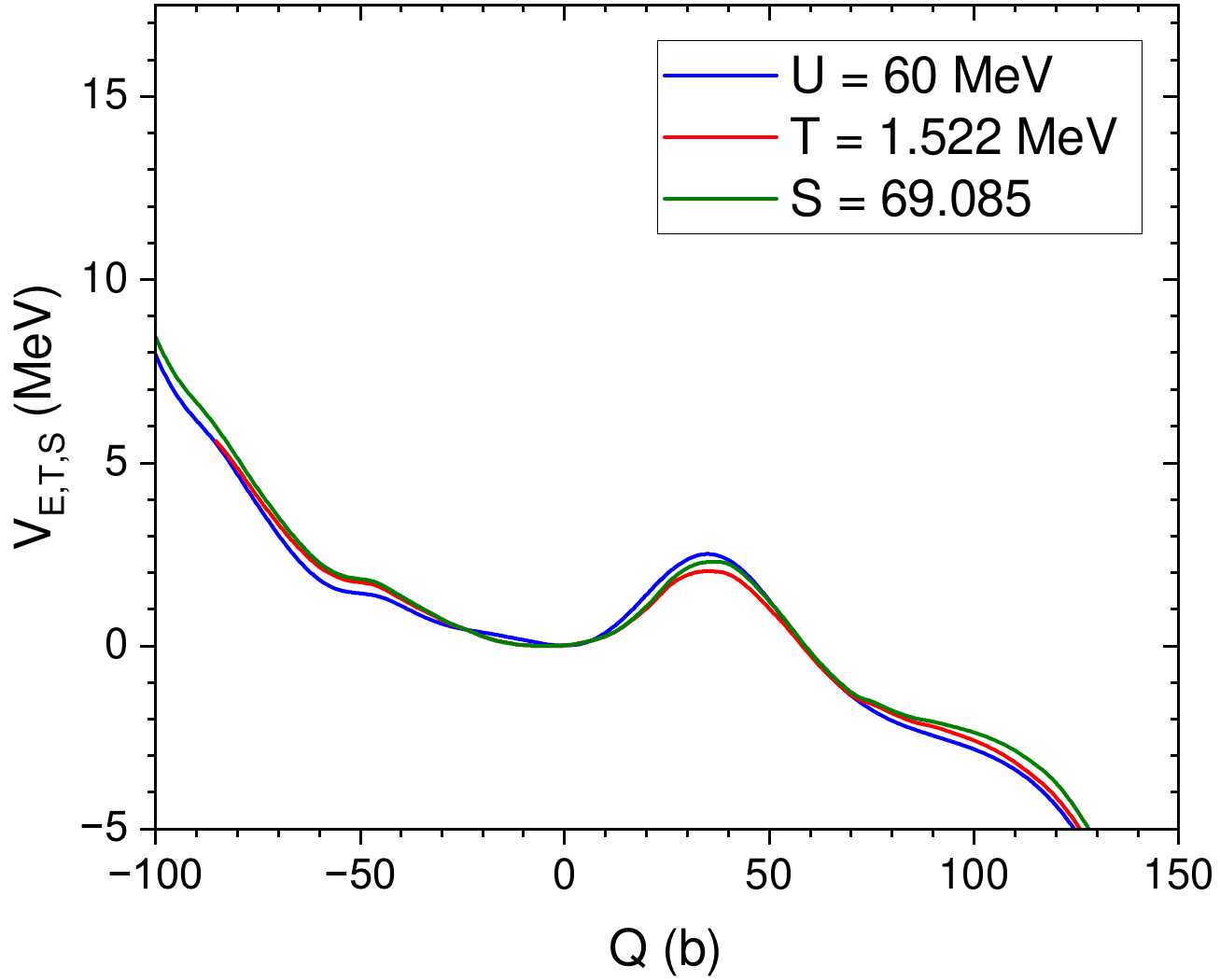}

    \caption{Calculated isoenergetic, isentropic  and isothermal effective  potentials for $^{296}$Lv. The minimum of the potential is normalized to zero in each case. In the isentropic (isothermal) description, the value $S$ ($T$) is fixed at the minimum of the isoenergetic potential $\Vpot{E}(Q)$ according to Eq.~\eqref{X_Qgs}. }
    \label{potentials}
\end{figure*}

Now we compare in Fig.~\ref{potentials} the effective potentials derived for three thermodynamic processes at the same excitation energies as in Fig.~\ref{potential_E}. Note  that at a given excitation energy $U$, the isentropic and isothermal effective potentials are obtained  by fixing the entropy $S$ and temperature $T$, respectively, to their values at the minimum of the corresponding isoenergetic potential.
Hence, for all three thermodynamic processes, the nucleus has the same total (and excitation) energy at the deformation~$\Qgs^*=0$
\begin{equation}\label{X_Qgs}
  E(\Qgs^*,X) = U + \Egs,~~~~(X=T,\,S,\,E).
  \end{equation}
With this definition of temperature and entropy, relations~\eqref{kappa_eq} ensure that, in all three thermodynamic processes, the deformation $\Qgs^*$ corresponds to a minimum of the respective effective potential.  Of course, this does not guarantee that the potentials $\Vpot{X}(Q)$ $(X=T,\,S,\,E)$ have exactly the same saddle point deformations. A common saddle point can be enforced by applying condition~\eqref{X_Qgs} at $Q=\Qsp$. However, in this case, since the total energy is not conserved in the isothermal and isentropic processes, the equilibrium minima of the effective potentials correspond to different values of the excitation energy $U$.
Thus, choosing the temperature and entropy according to~\eqref{X_Qgs} appears to be physically reasonable when comparing the effective potentials $\Vpot{X}(Q)$ $(X=T,\,S,\,E)$ at the same excitation energy.

Referring to Fig.~\ref{potentials}, in $^{296}$Lv all three effective potential curves achieve their saddle point at nearly the same deformation, $\Qsp\approx 39$\,b. As excitation energy grows, not only does the isoenergetic barrier diminish, but the isentropic and isothermal barriers also show gradual reduction. A crucial finding is that the barrier height $B_f$ is process dependent, attaining its maximum for the isoenergetic process. The largest discrepancies among $B_f(X=\text{const})$ ($X=T,\,S,\,E$) are observed at low excitation energies.
To elucidate these phenomena,  we examine the difference between isoenergetic and isentropic barrier heights, assuming the same saddle point in both processes:
\begin{eqnarray}\label{deltaEf_1}
  \Delta B_f&=& B_f(E'=\mathrm{const}) - B_f(S'=\mathrm{const})
  \notag \\
  &=&\int_{\Qgs^*}^{\Qsp}\left[\kappa(Q,E') - \kappa(Q,S')\right]dQ,
\end{eqnarray}
where $E'$ and $S'$ are related according to~\eqref{X_Qgs}.   Using Eqs.~\eqref{kappa_eq}, we obtain
\begin{eqnarray}\label{deltaEf_2}
  &&\Delta B_f = \int_{\Qgs^*}^{\Qsp}\left[\kappa(Q,E') - \kappa(Q,E(Q,S'))\right]dQ
    \notag\\
   &&\approx \int_{\Qgs^*}^{\Qsp} \frac{\partial \kappa(Q,E)}{\partial E}\Big|_{E=E'} \left[E'-E(Q,S')\right]dQ.
\end{eqnarray}
Since $E(Q,S')=\Vpot{S'}(Q) $ and $E'$ is the minimal value of $\Vpot{S'}(Q)$,  the expression in square brackets must be nonpositive. Moreover, referring to Fig.~\ref{kappa_E}, we find that
\begin{equation}
\frac{\partial\kappa(Q,E)}{\partial E}<0~~\mathrm{for}~~\Qgs^* < Q<\Qsp,
\end{equation}
which corresponds to diminishing of nuclear stability with increasing excitation energy.
Consequently, it follows that $B_f(E') > B_f(S')$. This result appears to be general, as the analytical relationships used in its derivation are not specific to a particular nucleus, and the position of the saddle point exhibits considerable stability with respect to excitation energy.
Analogously, we can conclude that $B_f(E') > B_f(T')$.

\begin{figure}[htbp]
\centering
\includegraphics[width=0.85\linewidth]{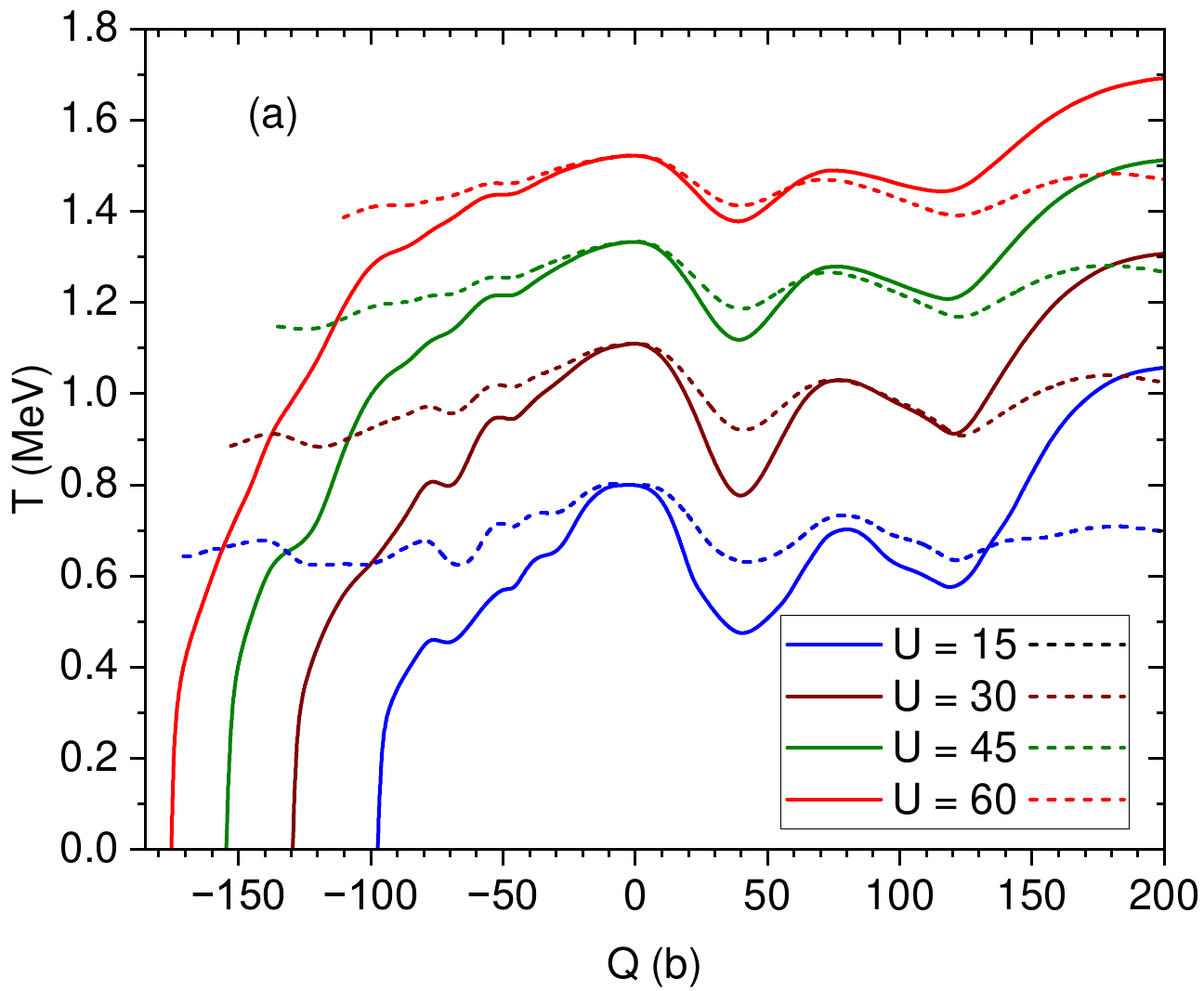}

\includegraphics[width=0.85\linewidth]{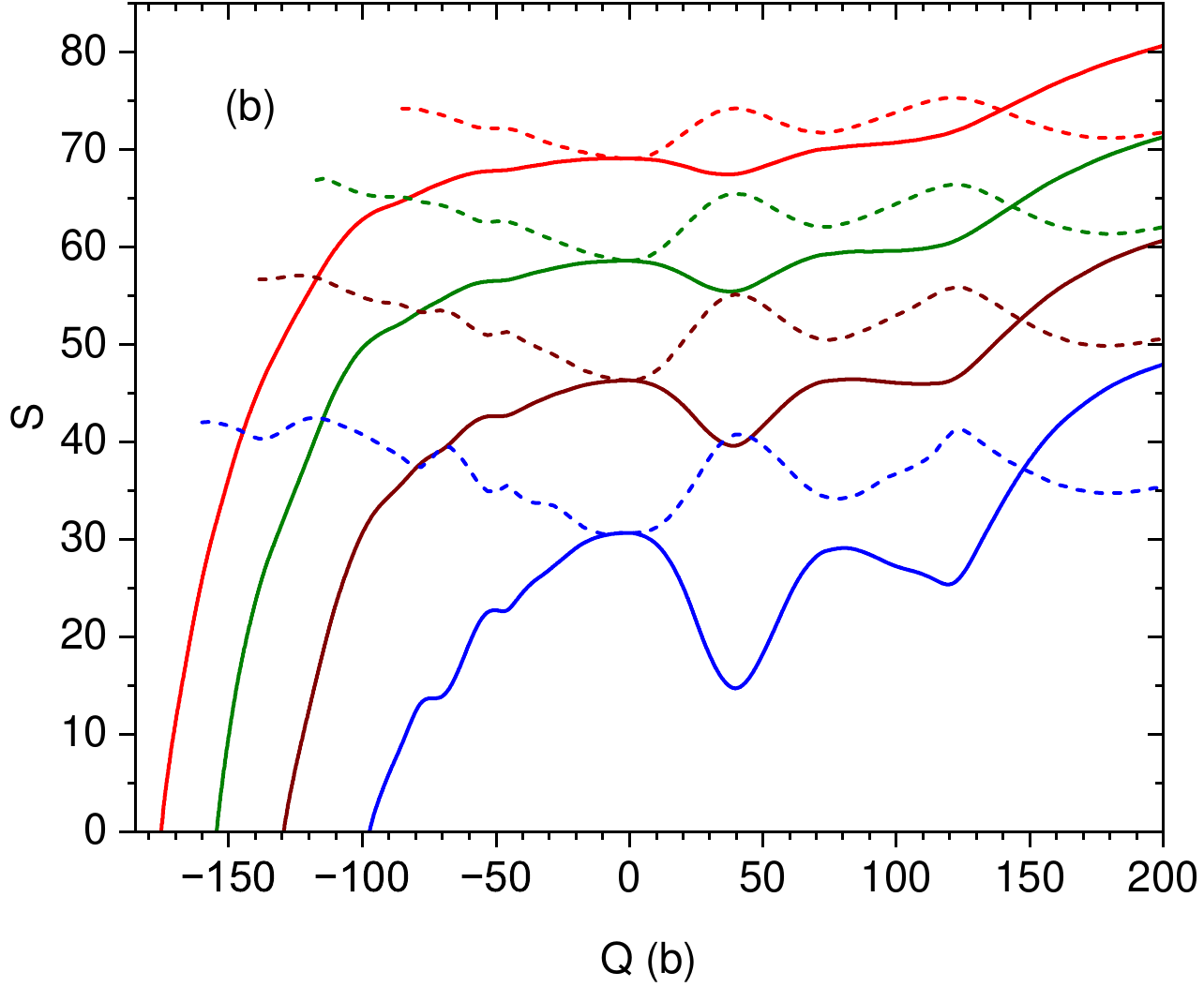}
    \caption{ (a) The temperature and (b) the entropy along the fission pathway of $^{296}$Lv at indicated excitation energies $U$ (in MeV). Solid lines correspond to the isoenergetic process, while dashed lines represent the temperature in the isentropic description and the entropy in the isothermal description.  The temperature $T$ (in the isentropic  scenario) and entropy $S$ (in the isothermal scenario) are displayed only within the deformation domain where the excitation energy does not exceed $70$\,MeV. }
    \label{temp_entr}
\end{figure}

The  evolution of temperature $T$ and  entropy $S$ along the fission pathway of $^{296}$Lv is shown in Fig.~\ref{temp_entr}.
It is seen that in the isoenergetic description both $T$ and $S$ change as a function of deformation. As it should be, both $T$ and $S$ vanish at the boundary $\Qcr$ when the entire excitation energy $U$ is completely converted to deformation energy.
As the nucleus moves away from the boundary, both $T$ and $S$ undergo a steep rise, with  the magnitude of growth intensifying with increasing excitation energy.
 The behavior of $T$ and $S$ directly reflects the effective potential landscape: they both attain maximal values at the potential minima and minimal values at saddle points. In addition, just like with the effective potential, higher excitation energy leads to a smaller difference between the maximum and minimum temperature and entropy values in the fission barrier region.

In Fig.~\ref{temp_entr}, we also compare the evolution of the temperature $T$ and entropy $S$ within the isoenergetic process with their behavior in the isothermal and isentropic descriptions. As seen in the figure, in the isoenergetic process, both $T$ and $S$ evolve along the fission pathway in a manner that is different from either the isothermal or isentropic scenarios.
The most striking difference emerges when comparing the isoenergetic and isothermal approaches: in the former, the entropy decreases as we approach the saddle point, whereas in the latter, the entropy increases. This contrasting behavior underscores the fundamentally different physical assumptions underlying these two thermodynamic descriptions.

\begin{figure}[htbp]
\centering
\includegraphics[width=0.85\columnwidth]{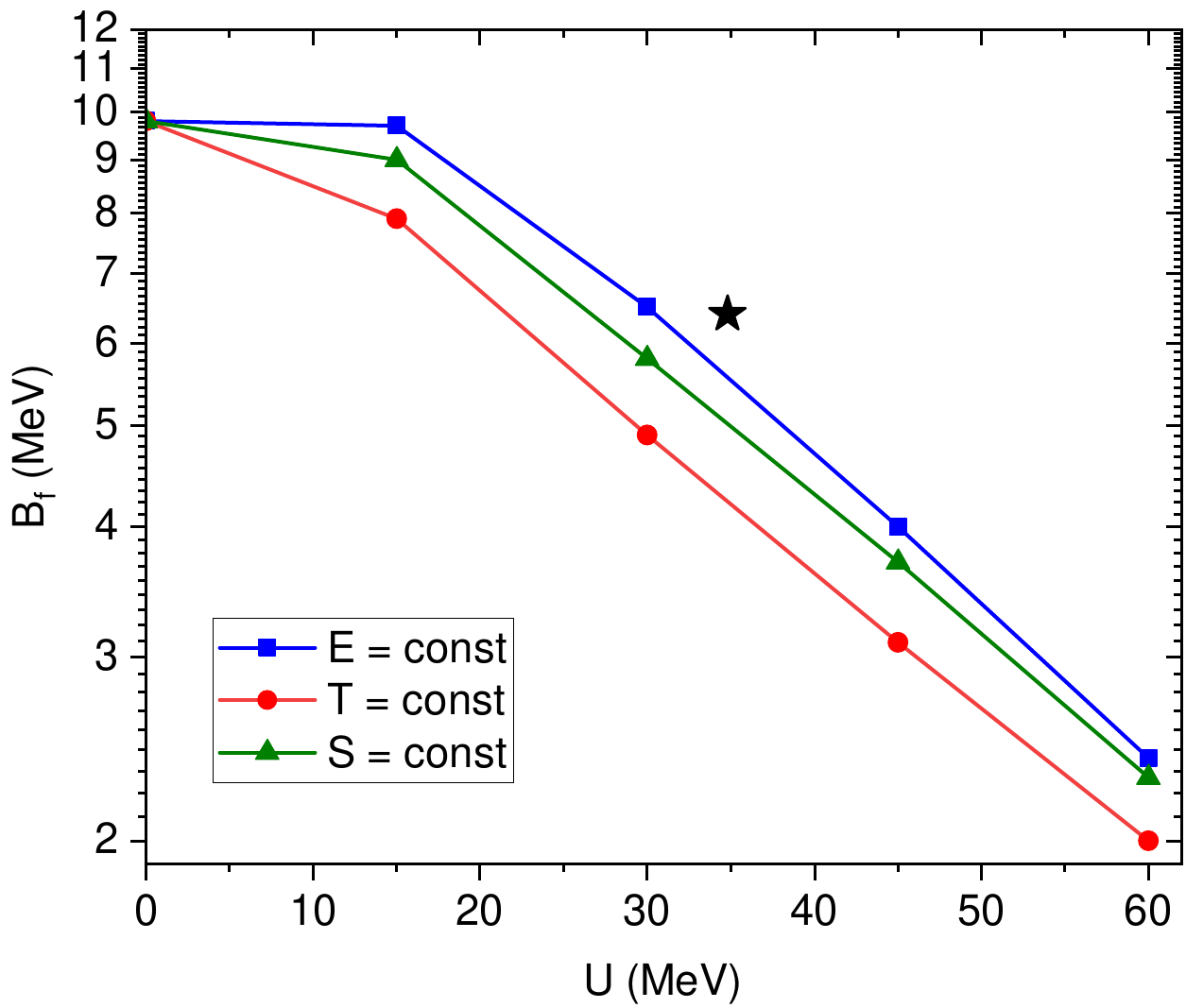}
    \caption{The heights of the fission barrier in $^{296}$Lv as a function of the excitation energy $U$.  The star denotes the lower limit of the fission barrier deduced from experimental data at excitation energy of 38.4 MeV~\cite{PRC65_Itkis}. }
    \label{barrier_height}
\end{figure}

\subsection{Damping of fission barrier height}

As shown in Fig.~\ref{potentials}, the difference $\Delta B_f$  between the fission barrier heights for the three thermodynamic processes decreases with increasing excitation energy, which is expected  for fission barriers approaching zero at very high excitation energies. The key question is whether the suppression rate of the fission barriers is universal or depends on the thermodynamic process.

In experimental analyses, the excitation energy dependence of fission barriers is usually parametrized by the phenomenological expression~\cite{PRC65_Itkis}
\begin{equation}\label{damping}
B_f \propto e^{-\gamma_D U},
\end{equation}
where the damping factor $\gamma_D$ quantifies the rate of barrier suppression with increasing excitation energy. To illustrate the differences in the damping factor among the three thermodynamic processes, in Fig.~\ref{barrier_height} we plot the fission barrier heights $B_f(X)$ as a function of $U$. The figure  clearly shows that  parametrization~\eqref{damping} provides an excellent description of the barrier heights at excitation energies above the ground-state barrier, allowing for a reliable extraction of the damping factor for each considered thermodynamic process. Extracting the slope of $\ln B_f$ we get $\gamma_D^{-1}(T)=33.5$\,MeV, $\gamma_D^{-1}(S)=32.4$\,MeV and $\gamma_D^{-1}(E)=30.1$\,MeV. Thus, among the three thermodynamic processes considered, the isoenergetic process is found to produce the maximum damping factor which is about two times larger than that
in the expression for the level-density parameter \cite{Ignatyuk_book}. It is worth noting that at low excitation energies, the suppression of pairing correlations can result in a small enhancement of the fission barriers~\cite{IJMPE18_Martin,PRC91_Schunck}.
In Ref.~\cite{PRC65_Itkis}, analysis of available experimental data for $^{296}$Lv yielded a lower limit of the fission barrier, $B_f \ge 6.4$\,MeV, at an excitation energy $U = 34.8$\,MeV.  Although the experimental lower limit of $B_f$ exceeds the barrier heights predicted by present microscopic calculations with SkM*, it shows better agreement with the prediction of the isoenergetic approach (see Fig.~\ref{barrier_height}).

\subsection{Level-density parameter}

\begin{figure}[htbp]
\centering
\includegraphics[width=0.85\columnwidth]{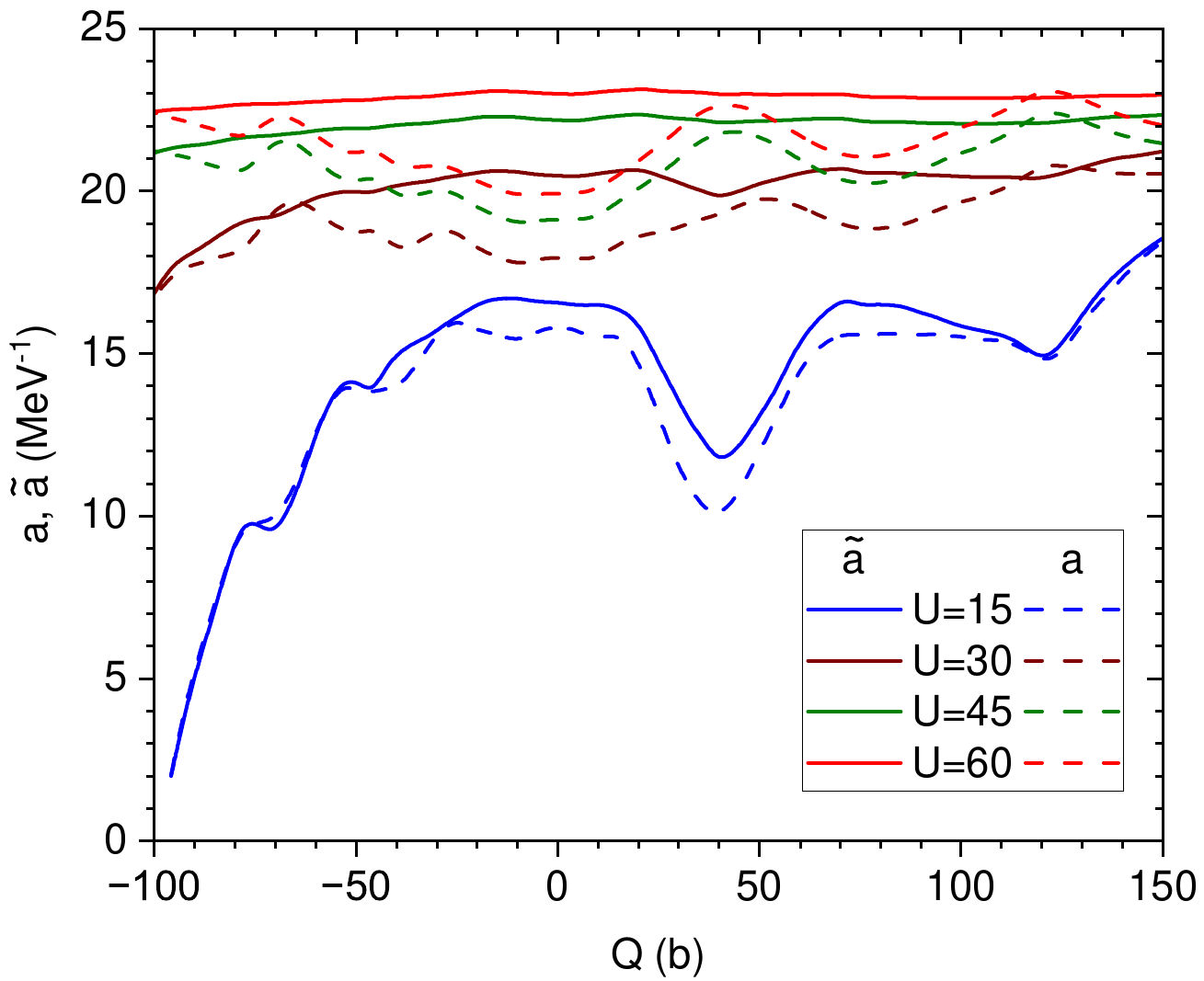}
    \caption{The level-density parameters $a$ and $\widetilde a$ (see the text for details) versus $Q$.}
    \label{level_dens_par}
\end{figure}

In Fig.~\ref{level_dens_par},  we examine how  the level-density parameter
\begin{equation}
  a(Q,E) = \frac{S^2(Q,E)}{4 U(Q)},~~~U(Q) = E-V(Q),
\end{equation}
varies with excitation energy and nuclear deformation. We observe that even at high excitation energies, the parameter shows a significant dependence on deformation. It is remarkable that the relationship between the level-density parameter  and the deformation energy landscape undergoes a significant transformation from low to high excitation energies.   At low excitation energies, the minima (maxima) of $a(Q)$ occur at the maxima (minima) of $V(Q)$, whereas at high excitation energies, the correspondence inverts: minima (maxima) of $a(Q)$ are found at the minima (maxima) of the deformation energy. This inversion is consistent with the results of micro-macro calculations for $^{296}$Lv~\cite{EPJA60_Rahmatinejad}, which predict that at low $U$, the value of $a(\Qsp)$ can be smaller than $a(\Qgs^*)$, and that the ratio $a(\Qsp)/a(\Qgs^*)$ increases with excitation energy. At high $U$ the present calculations gives $a(\Qsp)/a(\Qgs^*)\approx 1.18$, which is close to the asymptotic value $\approx 1.1$ predicted by the microscopic-macroscopic model~\cite{PRC105_Rahmatinejad}.

Finally, following~\cite{PRC88_Randrup}, we also consider an effective level-density parameter, which depends on the effective excitation energy
\begin{equation}
  \widetilde a(Q,E) = \frac{S^2(Q,E)}{4 U_E(Q)},~~U_E(Q) = E-\Vpot{E}(Q).
\end{equation}
As shown in Fig.~\ref{level_dens_par}, at low excitation energies, the effective level density parameter $\widetilde a(Q)$
behaves similarly to the  parameter $a(Q)$. However, as the excitation energy increases and the shell correction correspondingly decreases, the dependence of $\widetilde a(Q)$ on deformation becomes progressively smoother. Eventually, $\widetilde a(Q)$   becomes nearly independent of deformation, in accordence with the Fermi-gas model systematics~\cite{SovJNPhy29_Ignatyuk,Ignatyuk_book} . It is also worth noting that the difference between $\widetilde a(Q)$ and $a(Q)$
 is most pronounced in the regions of the minima of the effective potential, where the shell correction is large. In contrast, at the maxima of the effective potential, where the shell correction is small, the values of $\widetilde a(Q)$ and $a(Q)$
 become nearly identical.

\section{Conclusions}

In this work, the isoenergetic description of induced fission pathways in $^{296}$Lv was investigated within the framework of the EDFT. By employing self-consistent finite-temperature Hartree-Fock-Bogoliubov  calculations with the Skyrme-type interaction SkM$^*$, we derived and analyzed the effective potentials corresponding to different thermodynamic processes---isothermal, isentropic, and isoenergetic---focusing particularly on their dependence on nuclear deformation and excitation energy.

Our results confirm that the isoenergetic approach provides a physically consistent and relevant description for isolated excited nuclei, where total energy conservation plays a central role. The calculated driving force and associated effective potential reveal a significant reduction of the fission barrier with increasing excitation energy, in agreement with the expected thermal damping of shell effects. An important result of our study is the demonstration that the nonpotential rearrangement term plays a crucial role in shaping the isoenergetic effective potential.

Comparing the three different  thermodynamic processes, we showed that the respective effective potentials are not identical and the difference is most pronounced at  low and moderate excitation energies, where the isoenergetic approach predicts the highest fission barrier.
We also find that the isoenergetic process leads to the largest damping factor of the fission barrier compared to isothermal and isentropic descriptions.

Furthermore, we demonstrated that within the isoenergetic framework, the level-density parameter $a(Q)$ shows a pronounced dependence on nuclear deformation even at high excitation energies. In contrast, its effective counterpart $\widetilde{a}(Q)$ gradually approaches a deformation-independent Fermi-gas-like behavior as the excitation energy increases. Although other predictions of the Fermi-gas model, such as proportionality of the level-density parameter to the mass number, the exponential rise of level density with excitation energy, and so on, were  not  tested in the present study, the observed evolution of $\widetilde{a}(Q)$ reflects the progressive washing out of shell effects and provides some support for the applicability of the Fermi-gas model in the regime of high excitation energies.

The present study highlights the importance of using an appropriate thermodynamic framework when modeling fission pathways in excited nuclei and demonstrates the feasibility of incorporating isoenergetic conditions into the EDFT-based fission modeling. We considered only the one-dimensional fission pathway assuming axial and reflection symmetry. However,  the double-humped fission barrier in actinide nuclei is sensitive to the inclusion of nonaxial and reflection-asymmetric deformations~\cite{PRC80_Sheikh,PRC90_Schunck}. In this regard, the study of isoenergetic, symmetry-unrestricted, multidimensional fission pathways, as well as the relative suppression of the inner and outer fission barriers, is of great interest. In the future, we plan to continue  the study using the symmetry-unrestricted EDFT solvers, which will allow for a more comprehensive treatment of both nonaxial and reflection-asymmetric deformations.
The EDFT solver HFODD~\cite{CPC_180_Dobaczewski, CPC183_Schunck} is particularly well suited for such calculations, as it allows for the breaking of all self-consistent mean-field symmetries. Symmetry-unrestricted calculations of this type were previously carried out within both isothermal and isentropic frameworks~\cite{PRL102_Pei,PRC80_Sheikh,PRC87_McDonnell,PRC91_Schunck}.
We note, however,  that the main conclusion of our work---the progressive reduction and eventual disappearance of shell effects with increasing excitation energy---is consistent with the expected universal behavior in the high-temperature limit, as supported by both experimental data and theoretical models. Therefore, while the exact topology of the fission landscape may change in a symmetry-unrestricted treatment, the overall trend toward barrier suppression is expected to persist.

\acknowledgments
Useful discussions with Timur Shneidman and Azam Rahmatinejad are very gratefully acknowledged. This work was supported by Grant No. 25-42-00018 from
the Russian Science Foundation.


\begin{thebibliography}{46}%
\makeatletter
\providecommand \@ifxundefined [1]{%
 \@ifx{#1\undefined}
}%
\providecommand \@ifnum [1]{%
 \ifnum #1\expandafter \@firstoftwo
 \else \expandafter \@secondoftwo
 \fi
}%
\providecommand \@ifx [1]{%
 \ifx #1\expandafter \@firstoftwo
 \else \expandafter \@secondoftwo
 \fi
}%
\providecommand \natexlab [1]{#1}%
\providecommand \enquote  [1]{``#1''}%
\providecommand \bibnamefont  [1]{#1}%
\providecommand \bibfnamefont [1]{#1}%
\providecommand \citenamefont [1]{#1}%
\providecommand \href@noop [0]{\@secondoftwo}%
\providecommand \href [0]{\begingroup \@sanitize@url \@href}%
\providecommand \@href[1]{\@@startlink{#1}\@@href}%
\providecommand \@@href[1]{\endgroup#1\@@endlink}%
\providecommand \@sanitize@url [0]{\catcode `\\12\catcode `\$12\catcode
  `\&12\catcode `\#12\catcode `\^12\catcode `\_12\catcode `\%12\relax}%
\providecommand \@@startlink[1]{}%
\providecommand \@@endlink[0]{}%
\providecommand \url  [0]{\begingroup\@sanitize@url \@url }%
\providecommand \@url [1]{\endgroup\@href {#1}{\urlprefix }}%
\providecommand \urlprefix  [0]{URL }%
\providecommand \Eprint [0]{\href }%
\providecommand \doibase [0]{https://doi.org/}%
\providecommand \selectlanguage [0]{\@gobble}%
\providecommand \bibinfo  [0]{\@secondoftwo}%
\providecommand \bibfield  [0]{\@secondoftwo}%
\providecommand \translation [1]{[#1]}%
\providecommand \BibitemOpen [0]{}%
\providecommand \bibitemStop [0]{}%
\providecommand \bibitemNoStop [0]{.\EOS\space}%
\providecommand \EOS [0]{\spacefactor3000\relax}%
\providecommand \BibitemShut  [1]{\csname bibitem#1\endcsname}%
\let\auto@bib@innerbib\@empty
\bibitem [{\citenamefont {M\"oller}\ \emph {et~al.}(2009)\citenamefont
  {M\"oller}, \citenamefont {Sierk}, \citenamefont {Ichikawa}, \citenamefont
  {Iwamoto}, \citenamefont {Bengtsson}, \citenamefont {Uhrenholt},\ and\
  \citenamefont {\AA{}berg}}]{PRC79_Moller}%
  \BibitemOpen
  \bibfield  {author} {\bibinfo {author} {\bibfnamefont {P.}~\bibnamefont
  {M\"oller}}, \bibinfo {author} {\bibfnamefont {A.~J.}\ \bibnamefont {Sierk}},
  \bibinfo {author} {\bibfnamefont {T.}~\bibnamefont {Ichikawa}}, \bibinfo
  {author} {\bibfnamefont {A.}~\bibnamefont {Iwamoto}}, \bibinfo {author}
  {\bibfnamefont {R.}~\bibnamefont {Bengtsson}}, \bibinfo {author}
  {\bibfnamefont {H.}~\bibnamefont {Uhrenholt}},\ and\ \bibinfo {author}
  {\bibfnamefont {S.}~\bibnamefont {\AA{}berg}},\ }\bibfield  {title} {\bibinfo
  {title} {Heavy-element fission barriers},\ }\href
  {https://doi.org/10.1103/PhysRevC.79.064304} {\bibfield  {journal} {\bibinfo
  {journal} {Phys. Rev. C}\ }\textbf {\bibinfo {volume} {79}},\ \bibinfo
  {pages} {064304} (\bibinfo {year} {2009})}\BibitemShut {NoStop}%
\bibitem [{\citenamefont {Schunck}\ and\ \citenamefont
  {Robledo}(2016)}]{RepProgPhys81_Schunck}%
  \BibitemOpen
  \bibfield  {author} {\bibinfo {author} {\bibfnamefont {N.}~\bibnamefont
  {Schunck}}\ and\ \bibinfo {author} {\bibfnamefont {L.~M.}\ \bibnamefont
  {Robledo}},\ }\bibfield  {title} {\bibinfo {title} {Microscopic theory of
  nuclear fission: a review},\ }\href
  {https://doi.org/10.1088/0034-4885/79/11/116301} {\bibfield  {journal}
  {\bibinfo  {journal} {Reports on Progress in Physics}\ }\textbf {\bibinfo
  {volume} {79}},\ \bibinfo {pages} {116301} (\bibinfo {year}
  {2016})}\BibitemShut {NoStop}%
\bibitem [{\citenamefont {Andreyev}\ \emph {et~al.}(2017)\citenamefont
  {Andreyev}, \citenamefont {Nishio},\ and\ \citenamefont
  {Schmidt}}]{RepProgPhys81_Andreyev}%
  \BibitemOpen
  \bibfield  {author} {\bibinfo {author} {\bibfnamefont {A.~N.}\ \bibnamefont
  {Andreyev}}, \bibinfo {author} {\bibfnamefont {K.}~\bibnamefont {Nishio}},\
  and\ \bibinfo {author} {\bibfnamefont {K.-H.}\ \bibnamefont {Schmidt}},\
  }\bibfield  {title} {\bibinfo {title} {Nuclear fission: a review of
  experimental advances and phenomenology},\ }\href
  {https://doi.org/10.1088/1361-6633/aa82eb} {\bibfield  {journal} {\bibinfo
  {journal} {Reports on Progress in Physics}\ }\textbf {\bibinfo {volume}
  {81}},\ \bibinfo {pages} {016301} (\bibinfo {year} {2017})}\BibitemShut
  {NoStop}%
\bibitem [{\citenamefont {Schmidt}\ and\ \citenamefont
  {Jurado}(2018)}]{RepProgPhys81_Schmidt}%
  \BibitemOpen
  \bibfield  {author} {\bibinfo {author} {\bibfnamefont {K.-H.}\ \bibnamefont
  {Schmidt}}\ and\ \bibinfo {author} {\bibfnamefont {B.}~\bibnamefont
  {Jurado}},\ }\bibfield  {title} {\bibinfo {title} {Review on the progress in
  nuclear fission—experimental methods and theoretical descriptions},\ }\href
  {https://doi.org/10.1088/1361-6633/aacfa7} {\bibfield  {journal} {\bibinfo
  {journal} {Reports on Progress in Physics}\ }\textbf {\bibinfo {volume}
  {81}},\ \bibinfo {pages} {106301} (\bibinfo {year} {2018})}\BibitemShut
  {NoStop}%
\bibitem [{\citenamefont {Jachimowicz}\ \emph {et~al.}(2021)\citenamefont
  {Jachimowicz}, \citenamefont {Kowal},\ and\ \citenamefont
  {Skalski}}]{ADNDT138_Jachimowicz}%
  \BibitemOpen
  \bibfield  {author} {\bibinfo {author} {\bibfnamefont {P.}~\bibnamefont
  {Jachimowicz}}, \bibinfo {author} {\bibfnamefont {M.}~\bibnamefont {Kowal}},\
  and\ \bibinfo {author} {\bibfnamefont {J.}~\bibnamefont {Skalski}},\
  }\bibfield  {title} {\bibinfo {title} {Properties of heaviest nuclei with
  98$\le${Z}$\le$126 and 134$\le$ {N}$\le$192},\ }\href
  {https://doi.org/https://doi.org/10.1016/j.adt.2020.101393} {\bibfield
  {journal} {\bibinfo  {journal} {Atomic Data and Nuclear Data Tables}\
  }\textbf {\bibinfo {volume} {138}},\ \bibinfo {pages} {101393} (\bibinfo
  {year} {2021})}\BibitemShut {NoStop}%
\bibitem [{\citenamefont {Schunck}(2023)}]{Schunck2023}%
  \BibitemOpen
  \bibfield  {author} {\bibinfo {author} {\bibfnamefont {N.}~\bibnamefont
  {Schunck}},\ }\bibinfo {title} {Microscopic theory of nuclear fission},\ in\
  \href {https://doi.org/10.1007/978-981-19-6345-2_80} {\emph {\bibinfo
  {booktitle} {Handbook of Nuclear Physics}}},\ \bibinfo {editor} {edited by\
  \bibinfo {editor} {\bibfnamefont {I.}~\bibnamefont {Tanihata}}, \bibinfo
  {editor} {\bibfnamefont {H.}~\bibnamefont {Toki}},\ and\ \bibinfo {editor}
  {\bibfnamefont {T.}~\bibnamefont {Kajino}}}\ (\bibinfo  {publisher} {Springer
  Nature Singapore},\ \bibinfo {address} {Singapore},\ \bibinfo {year} {2023})\
  pp.\ \bibinfo {pages} {829--866}\BibitemShut {NoStop}%
\bibitem [{\citenamefont {Itkis}\ \emph {et~al.}(2002)\citenamefont {Itkis},
  \citenamefont {Oganessian},\ and\ \citenamefont {Zagrebaev}}]{PRC65_Itkis}%
  \BibitemOpen
  \bibfield  {author} {\bibinfo {author} {\bibfnamefont {M.~G.}\ \bibnamefont
  {Itkis}}, \bibinfo {author} {\bibfnamefont {Y.~T.}\ \bibnamefont
  {Oganessian}},\ and\ \bibinfo {author} {\bibfnamefont {V.~I.}\ \bibnamefont
  {Zagrebaev}},\ }\bibfield  {title} {\bibinfo {title} {Fission barriers of
  superheavy nuclei},\ }\href {https://doi.org/10.1103/PhysRevC.65.044602}
  {\bibfield  {journal} {\bibinfo  {journal} {Phys. Rev. C}\ }\textbf {\bibinfo
  {volume} {65}},\ \bibinfo {pages} {044602} (\bibinfo {year}
  {2002})}\BibitemShut {NoStop}%
\bibitem [{\citenamefont {Loveland}(2007)}]{PRC76_Loveland}%
  \BibitemOpen
  \bibfield  {author} {\bibinfo {author} {\bibfnamefont {W.}~\bibnamefont
  {Loveland}},\ }\bibfield  {title} {\bibinfo {title} {Synthesis of
  transactinide nuclei using radioactive beams},\ }\href
  {https://doi.org/10.1103/PhysRevC.76.014612} {\bibfield  {journal} {\bibinfo
  {journal} {Phys. Rev. C}\ }\textbf {\bibinfo {volume} {76}},\ \bibinfo
  {pages} {014612} (\bibinfo {year} {2007})}\BibitemShut {NoStop}%
\bibitem [{\citenamefont {Rahmatinejad}\ \emph {et~al.}(2024)\citenamefont
  {Rahmatinejad}, \citenamefont {Shneidman}, \citenamefont {Adamian},
  \citenamefont {Antonenko}, \citenamefont {Jachimowicz},\ and\ \citenamefont
  {Kowal}}]{EPJA60_Rahmatinejad}%
  \BibitemOpen
  \bibfield  {author} {\bibinfo {author} {\bibfnamefont {A.}~\bibnamefont
  {Rahmatinejad}}, \bibinfo {author} {\bibfnamefont {T.~M.}\ \bibnamefont
  {Shneidman}}, \bibinfo {author} {\bibfnamefont {G.~G.}\ \bibnamefont
  {Adamian}}, \bibinfo {author} {\bibfnamefont {N.~V.}\ \bibnamefont
  {Antonenko}}, \bibinfo {author} {\bibfnamefont {P.}~\bibnamefont
  {Jachimowicz}},\ and\ \bibinfo {author} {\bibfnamefont {M.}~\bibnamefont
  {Kowal}},\ }\bibfield  {title} {\bibinfo {title} {Entropies, level-density
  parameters and fission probabilities along the triaxially- and
  axially-symmetric fission paths in $^{296}${Lv}},\ }\href
  {https://doi.org/10.1140/epja/s10050-024-01437-w} {\bibfield  {journal}
  {\bibinfo  {journal} {The European Physical Journal A}\ }\textbf {\bibinfo
  {volume} {60}},\ \bibinfo {pages} {214} (\bibinfo {year} {2024})}\BibitemShut
  {NoStop}%
\bibitem [{\citenamefont {Sauer}\ \emph {et~al.}(1976)\citenamefont {Sauer},
  \citenamefont {Chandra},\ and\ \citenamefont {Mosel}}]{NPA264_Sauer}%
  \BibitemOpen
  \bibfield  {author} {\bibinfo {author} {\bibfnamefont {G.}~\bibnamefont
  {Sauer}}, \bibinfo {author} {\bibfnamefont {H.}~\bibnamefont {Chandra}},\
  and\ \bibinfo {author} {\bibfnamefont {U.}~\bibnamefont {Mosel}},\ }\bibfield
   {title} {\bibinfo {title} {Thermal properties of nuclei},\ }\href
  {https://doi.org/https://doi.org/10.1016/0375-9474(76)90429-2} {\bibfield
  {journal} {\bibinfo  {journal} {Nuclear Physics A}\ }\textbf {\bibinfo
  {volume} {264}},\ \bibinfo {pages} {221} (\bibinfo {year}
  {1976})}\BibitemShut {NoStop}%
\bibitem [{\citenamefont {Okolowicz}\ and\ \citenamefont
  {Irvine}(1987)}]{JPhG13_Okolowicz}%
  \BibitemOpen
  \bibfield  {author} {\bibinfo {author} {\bibfnamefont {J.}~\bibnamefont
  {Okolowicz}}\ and\ \bibinfo {author} {\bibfnamefont {J.~M.}\ \bibnamefont
  {Irvine}},\ }\bibfield  {title} {\bibinfo {title} {The temperature dependence
  of the fission barrier in a mean-field description of rotating nuclei},\
  }\href {https://doi.org/10.1088/0305-4616/13/11/012} {\bibfield  {journal}
  {\bibinfo  {journal} {Journal of Physics G: Nuclear Physics}\ }\textbf
  {\bibinfo {volume} {13}},\ \bibinfo {pages} {1399} (\bibinfo {year}
  {1987})}\BibitemShut {NoStop}%
\bibitem [{\citenamefont {Martin}\ and\ \citenamefont
  {Robledo}(2009)}]{IJMPE18_Martin}%
  \BibitemOpen
  \bibfield  {author} {\bibinfo {author} {\bibfnamefont {V.}~\bibnamefont
  {Martin}}\ and\ \bibinfo {author} {\bibfnamefont {L.~M.}\ \bibnamefont
  {Robledo}},\ }\bibfield  {title} {\bibinfo {title} {Fission barriers at
  finite temperature: a theoretical description with the {G}ogny force},\
  }\href {https://doi.org/10.1142/S0218301309012963} {\bibfield  {journal}
  {\bibinfo  {journal} {International Journal of Modern Physics E}\ }\textbf
  {\bibinfo {volume} {18}},\ \bibinfo {pages} {861} (\bibinfo {year}
  {2009})}\BibitemShut {NoStop}%
\bibitem [{\citenamefont {Schunck}\ \emph {et~al.}(2015)\citenamefont
  {Schunck}, \citenamefont {Duke},\ and\ \citenamefont {Carr}}]{PRC91_Schunck}%
  \BibitemOpen
  \bibfield  {author} {\bibinfo {author} {\bibfnamefont {N.}~\bibnamefont
  {Schunck}}, \bibinfo {author} {\bibfnamefont {D.}~\bibnamefont {Duke}},\ and\
  \bibinfo {author} {\bibfnamefont {H.}~\bibnamefont {Carr}},\ }\bibfield
  {title} {\bibinfo {title} {Description of induced nuclear fission with
  {Skyrme} energy functionals. {II.} {Finite} temperature effects},\ }\href
  {https://doi.org/10.1103/PhysRevC.91.034327} {\bibfield  {journal} {\bibinfo
  {journal} {Phys. Rev. C}\ }\textbf {\bibinfo {volume} {91}},\ \bibinfo
  {pages} {034327} (\bibinfo {year} {2015})}\BibitemShut {NoStop}%
\bibitem [{\citenamefont
  {Schunck}(2013{\natexlab{a}})}]{JPhConfSer436_Schunck}%
  \BibitemOpen
  \bibfield  {author} {\bibinfo {author} {\bibfnamefont {N.}~\bibnamefont
  {Schunck}},\ }\bibfield  {title} {\bibinfo {title} {Microscopic description
  of induced fission},\ }\href {https://doi.org/10.1088/1742-6596/436/1/012058}
  {\bibfield  {journal} {\bibinfo  {journal} {Journal of Physics: Conference
  Series}\ }\textbf {\bibinfo {volume} {436}},\ \bibinfo {pages} {012058}
  (\bibinfo {year} {2013}{\natexlab{a}})}\BibitemShut {NoStop}%
\bibitem [{\citenamefont {Goodman}(1981)}]{Goodman_NPA352}%
  \BibitemOpen
  \bibfield  {author} {\bibinfo {author} {\bibfnamefont {A.~L.}\ \bibnamefont
  {Goodman}},\ }\bibfield  {title} {\bibinfo {title} {Finite-temperature {HFB}
  theory},\ }\href
  {https://doi.org/https://doi.org/10.1016/0375-9474(81)90557-1} {\bibfield
  {journal} {\bibinfo  {journal} {Nuclear Physics A}\ }\textbf {\bibinfo
  {volume} {352}},\ \bibinfo {pages} {30} (\bibinfo {year} {1981})}\BibitemShut
  {NoStop}%
\bibitem [{\citenamefont {Schunck}(2019)}]{Schunck_book}%
  \BibitemOpen
  \bibfield  {author} {\bibinfo {author} {\bibfnamefont {N.}~\bibnamefont
  {Schunck}},\ }\href {https://doi.org/10.1088/2053-2563/aae0ed} {\emph
  {\bibinfo {title} {Energy Density Functional Methods for Atomic Nuclei}}}\
  (\bibinfo  {publisher} {IOP Publishing, Bristol, U.K.},\ \bibinfo {year} {2019})\BibitemShut
  {NoStop}%
\bibitem [{\citenamefont {Dzhioev}\ and\ \citenamefont
  {Vdovin}(2022{\natexlab{a}})}]{PPN53_1_Dzhioev}%
  \BibitemOpen
  \bibfield  {author} {\bibinfo {author} {\bibfnamefont {A.~A.}\ \bibnamefont
  {Dzhioev}}\ and\ \bibinfo {author} {\bibfnamefont {A.~I.}\ \bibnamefont
  {Vdovin}},\ }\bibfield  {title} {\bibinfo {title} {Superoperator approach to
  the theory of hot nuclei and astrophysical applications: {I}. {S}pectral
  properties of hot nuclei},\ }\href
  {https://doi.org/10.1134/S1063779622050033} {\bibfield  {journal} {\bibinfo
  {journal} {Phys. Part. Nucl.}\ }\textbf {\bibinfo {volume}
  {53}},\ \bibinfo {pages} {885} (\bibinfo {year}
  {2022}{\natexlab{a}})}\BibitemShut {NoStop}%
\bibitem [{\citenamefont {Dzhioev}\ and\ \citenamefont
  {Vdovin}(2022{\natexlab{b}})}]{PPN53_2_Dzhioev}%
  \BibitemOpen
  \bibfield  {author} {\bibinfo {author} {\bibfnamefont {A.~A.}\ \bibnamefont
  {Dzhioev}}\ and\ \bibinfo {author} {\bibfnamefont {A.~I.}\ \bibnamefont
  {Vdovin}},\ }\bibfield  {title} {\bibinfo {title} {Superoperator approach to
  the theory of hot nuclei and astrophysical applications: {II}. {E}lectron
  capture in stars},\ }\href {https://doi.org/10.1134/S1063779622050045}
  {\bibfield  {journal} {\bibinfo  {journal} {Phys. Part. Nucl.}\
  }\textbf {\bibinfo {volume} {53}},\ \bibinfo {pages} {939} (\bibinfo {year}
  {2022}{\natexlab{b}})}\BibitemShut {NoStop}%
\bibitem [{\citenamefont {Dzhioev}\ and\ \citenamefont
  {Vdovin}(2022{\natexlab{c}})}]{PPN53_3_Dzhioev}%
  \BibitemOpen
  \bibfield  {author} {\bibinfo {author} {\bibfnamefont {A.~A.}\ \bibnamefont
  {Dzhioev}}\ and\ \bibinfo {author} {\bibfnamefont {A.~I.}\ \bibnamefont
  {Vdovin}},\ }\bibfield  {title} {\bibinfo {title} {Superoperator approach to
  the theory of hot nuclei and astrophysical applications: {III}.
  {N}eutrino--nucleus reactions in stars},\ }\href
  {https://doi.org/10.1134/S106377962206003X} {\bibfield  {journal} {\bibinfo
  {journal} {Phys. Part. Nucl.}\ }\textbf {\bibinfo {volume}
  {53}},\ \bibinfo {pages} {1051} (\bibinfo {year}
  {2022}{\natexlab{c}})}\BibitemShut {NoStop}%
\bibitem [{\citenamefont {Diebel}\ \emph {et~al.}(1981)\citenamefont {Diebel},
  \citenamefont {Albrecht},\ and\ \citenamefont {Hasse}}]{NPA335_Duebel}%
  \BibitemOpen
  \bibfield  {author} {\bibinfo {author} {\bibfnamefont {M.}~\bibnamefont
  {Diebel}}, \bibinfo {author} {\bibfnamefont {K.}~\bibnamefont {Albrecht}},\
  and\ \bibinfo {author} {\bibfnamefont {R.~W.}\ \bibnamefont {Hasse}},\
  }\bibfield  {title} {\bibinfo {title} {Microscopic calculations of fission
  barriers and critical angular momenta for excited heavy nuclear systems},\
  }\href {https://doi.org/https://doi.org/10.1016/0375-9474(81)90132-9}
  {\bibfield  {journal} {\bibinfo  {journal} {Nuclear Physics A}\ }\textbf
  {\bibinfo {volume} {355}},\ \bibinfo {pages} {66} (\bibinfo {year}
  {1981})}\BibitemShut {NoStop}%
\bibitem [{\citenamefont {Pei}\ \emph {et~al.}(2009)\citenamefont {Pei},
  \citenamefont {Nazarewicz}, \citenamefont {Sheikh},\ and\ \citenamefont
  {Kerman}}]{PRL102_Pei}%
  \BibitemOpen
  \bibfield  {author} {\bibinfo {author} {\bibfnamefont {J.~C.}\ \bibnamefont
  {Pei}}, \bibinfo {author} {\bibfnamefont {W.}~\bibnamefont {Nazarewicz}},
  \bibinfo {author} {\bibfnamefont {J.~A.}\ \bibnamefont {Sheikh}},\ and\
  \bibinfo {author} {\bibfnamefont {A.~K.}\ \bibnamefont {Kerman}},\ }\bibfield
   {title} {\bibinfo {title} {Fission barriers of compound superheavy nuclei},\
  }\href {https://doi.org/10.1103/PhysRevLett.102.192501} {\bibfield  {journal}
  {\bibinfo  {journal} {Phys. Rev. Lett.}\ }\textbf {\bibinfo {volume} {102}},\
  \bibinfo {pages} {192501} (\bibinfo {year} {2009})}\BibitemShut {NoStop}%
\bibitem [{\citenamefont {Sheikh}\ \emph {et~al.}(2009)\citenamefont {Sheikh},
  \citenamefont {Nazarewicz},\ and\ \citenamefont {Pei}}]{PRC80_Sheikh}%
  \BibitemOpen
  \bibfield  {author} {\bibinfo {author} {\bibfnamefont {J.~A.}\ \bibnamefont
  {Sheikh}}, \bibinfo {author} {\bibfnamefont {W.}~\bibnamefont {Nazarewicz}},\
  and\ \bibinfo {author} {\bibfnamefont {J.~C.}\ \bibnamefont {Pei}},\
  }\bibfield  {title} {\bibinfo {title} {Systematic study of fission barriers
  of excited superheavy nuclei},\ }\href
  {https://doi.org/10.1103/PhysRevC.80.011302} {\bibfield  {journal} {\bibinfo
  {journal} {Phys. Rev. C}\ }\textbf {\bibinfo {volume} {80}},\ \bibinfo
  {pages} {011302} (\bibinfo {year} {2009})}\BibitemShut {NoStop}%
\bibitem [{\citenamefont {Pei}\ \emph {et~al.}(2010)\citenamefont {Pei},
  \citenamefont {Nazarewicz}, \citenamefont {Sheikh},\ and\ \citenamefont
  {Kerman}}]{NPA834_Pei}%
  \BibitemOpen
  \bibfield  {author} {\bibinfo {author} {\bibfnamefont {J.}~\bibnamefont
  {Pei}}, \bibinfo {author} {\bibfnamefont {W.}~\bibnamefont {Nazarewicz}},
  \bibinfo {author} {\bibfnamefont {J.}~\bibnamefont {Sheikh}},\ and\ \bibinfo
  {author} {\bibfnamefont {A.}~\bibnamefont {Kerman}},\ }\bibfield  {title}
  {\bibinfo {title} {Fission barriers and neutron gas in compound superheavy
  nuclei},\ }\href
  {https://doi.org/https://doi.org/10.1016/j.nuclphysa.2010.01.045} {\bibfield
  {journal} {\bibinfo  {journal} {Nuclear Physics A}\ }\textbf {\bibinfo
  {volume} {834}},\ \bibinfo {pages} {381c} (\bibinfo {year}
  {2010})}\BibitemShut {NoStop}%
\bibitem [{\citenamefont {McDonnell}\ \emph {et~al.}(2013)\citenamefont
  {McDonnell}, \citenamefont {Nazarewicz},\ and\ \citenamefont
  {Sheikh}}]{PRC87_McDonnell}%
  \BibitemOpen
  \bibfield  {author} {\bibinfo {author} {\bibfnamefont {J.~D.}\ \bibnamefont
  {McDonnell}}, \bibinfo {author} {\bibfnamefont {W.}~\bibnamefont
  {Nazarewicz}},\ and\ \bibinfo {author} {\bibfnamefont {J.~A.}\ \bibnamefont
  {Sheikh}},\ }\bibfield  {title} {\bibinfo {title} {Third minima in thorium
  and uranium isotopes in a self-consistent theory},\ }\href
  {https://doi.org/10.1103/PhysRevC.87.054327} {\bibfield  {journal} {\bibinfo
  {journal} {Phys. Rev. C}\ }\textbf {\bibinfo {volume} {87}},\ \bibinfo
  {pages} {054327} (\bibinfo {year} {2013})}\BibitemShut {NoStop}%
\bibitem [{\citenamefont {Randrup}\ and\ \citenamefont
  {M\"oller}(2013)}]{PRC88_Randrup}%
  \BibitemOpen
  \bibfield  {author} {\bibinfo {author} {\bibfnamefont {J.}~\bibnamefont
  {Randrup}}\ and\ \bibinfo {author} {\bibfnamefont {P.}~\bibnamefont
  {M\"oller}},\ }\bibfield  {title} {\bibinfo {title} {Energy dependence of
  fission-fragment mass distributions from strongly damped shape evolution},\
  }\href {https://doi.org/10.1103/PhysRevC.88.064606} {\bibfield  {journal}
  {\bibinfo  {journal} {Phys. Rev. C}\ }\textbf {\bibinfo {volume} {88}},\
  \bibinfo {pages} {064606} (\bibinfo {year} {2013})}\BibitemShut {NoStop}%
\bibitem [{\citenamefont {Fröbrich}\ \emph {et~al.}(1993)\citenamefont
  {Fröbrich}, \citenamefont {Gontchar},\ and\ \citenamefont
  {Mavlitov}}]{NPA556_Frobrich}%
  \BibitemOpen
  \bibfield  {author} {\bibinfo {author} {\bibfnamefont {P.}~\bibnamefont
  {Fröbrich}}, \bibinfo {author} {\bibfnamefont {I.}~\bibnamefont
  {Gontchar}},\ and\ \bibinfo {author} {\bibfnamefont {N.}~\bibnamefont
  {Mavlitov}},\ }\bibfield  {title} {\bibinfo {title} {Langevin
  fluctuation-dissipation dynamics of hot nuclei: Prescission neutron
  multiplicities and fission probabilities},\ }\href
  {https://doi.org/https://doi.org/10.1016/0375-9474(93)90352-X} {\bibfield
  {journal} {\bibinfo  {journal} {Nuclear Physics A}\ }\textbf {\bibinfo
  {volume} {556}},\ \bibinfo {pages} {281} (\bibinfo {year}
  {1993})}\BibitemShut {NoStop}%
\bibitem [{\citenamefont {Fröbrich}\ and\ \citenamefont
  {Gontchar}(1998)}]{PhysRep292_Frobrich}%
  \BibitemOpen
  \bibfield  {author} {\bibinfo {author} {\bibfnamefont {P.}~\bibnamefont
  {Fröbrich}}\ and\ \bibinfo {author} {\bibfnamefont {I.}~\bibnamefont
  {Gontchar}},\ }\bibfield  {title} {\bibinfo {title} {Langevin description of
  fusion, deep-inelastic collisions and heavy-ion-induced fission},\ }\href
  {https://doi.org/https://doi.org/10.1016/S0370-1573(97)00042-2} {\bibfield
  {journal} {\bibinfo  {journal} {Physics Reports}\ }\textbf {\bibinfo {volume}
  {292}},\ \bibinfo {pages} {131} (\bibinfo {year} {1998})}\BibitemShut
  {NoStop}%
\bibitem [{\citenamefont {Schunck}(2013{\natexlab{b}})}]{APhPolB44_Schunck}%
  \BibitemOpen
  \bibfield  {author} {\bibinfo {author} {\bibfnamefont {N.}~\bibnamefont
  {Schunck}},\ }\bibfield  {title} {\bibinfo {title} {Density functional theory
  approach to nuclear fission},\ }\href
  {https://doi.org/https://doi.org/10.5506/APhysPolB.44.263} {\bibfield
  {journal} {\bibinfo  {journal} {Acta Physica Polonica B}\ }\textbf {\bibinfo
  {volume} {44}},\ \bibinfo {pages} {263} (\bibinfo {year}
  {2013}{\natexlab{b}})}\BibitemShut {NoStop}%
\bibitem [{\citenamefont {Balian}(2007)}]{Balian_book}%
  \BibitemOpen
  \bibfield  {author} {\bibinfo {author} {\bibfnamefont {R.}~\bibnamefont
  {Balian}},\ }\href {https://link.springer.com/book/9783540454694} {\emph
  {\bibinfo {title} {From Microphysics to Macrophysics: Methods and
  Applications of Statistical Physics}}},\ Vol.~\bibinfo {volume} {1}\
  (\bibinfo  {publisher} {Springer},\ \bibinfo {year} {2007})\BibitemShut
  {NoStop}%
\bibitem [{\citenamefont {Alhassid}\ \emph {et~al.}(2016)\citenamefont
  {Alhassid}, \citenamefont {Bertsch}, \citenamefont {Gilbreth},\ and\
  \citenamefont {Nakada}}]{PRC93_Alhassid}%
  \BibitemOpen
  \bibfield  {author} {\bibinfo {author} {\bibfnamefont {Y.}~\bibnamefont
  {Alhassid}}, \bibinfo {author} {\bibfnamefont {G.~F.}\ \bibnamefont
  {Bertsch}}, \bibinfo {author} {\bibfnamefont {C.~N.}\ \bibnamefont
  {Gilbreth}},\ and\ \bibinfo {author} {\bibfnamefont {H.}~\bibnamefont
  {Nakada}},\ }\bibfield  {title} {\bibinfo {title} {Benchmarking mean-field
  approximations to level densities},\ }\href
  {https://doi.org/10.1103/PhysRevC.93.044320} {\bibfield  {journal} {\bibinfo
  {journal} {Phys. Rev. C}\ }\textbf {\bibinfo {volume} {93}},\ \bibinfo
  {pages} {044320} (\bibinfo {year} {2016})}\BibitemShut {NoStop}%
\bibitem [{\citenamefont {Ryssens}\ and\ \citenamefont
  {Alhassid}(2021)}]{EPJA57_Ryssens}%
  \BibitemOpen
  \bibfield  {author} {\bibinfo {author} {\bibfnamefont {W.}~\bibnamefont
  {Ryssens}}\ and\ \bibinfo {author} {\bibfnamefont {Y.}~\bibnamefont
  {Alhassid}},\ }\bibfield  {title} {\bibinfo {title} {Finite-temperature
  mean-field approximations for shell model {Hamiltonians}: the code
  {HF-SHELL}},\ }\href {https://doi.org/10.1140/epja/s10050-021-00365-3}
  {\bibfield  {journal} {\bibinfo  {journal} {Eur. Phys. J. A}\ }\textbf
  {\bibinfo {volume} {57}},\ \bibinfo {pages} {76} (\bibinfo {year}
  {2021})}\BibitemShut {NoStop}%
\bibitem [{\citenamefont {Egido}\ and\ \citenamefont
  {Ring}(1993)}]{JPhG19_Egido}%
  \BibitemOpen
  \bibfield  {author} {\bibinfo {author} {\bibfnamefont {J.~L.}\ \bibnamefont
  {Egido}}\ and\ \bibinfo {author} {\bibfnamefont {P.}~\bibnamefont {Ring}},\
  }\bibfield  {title} {\bibinfo {title} {The decay of hot nuclei},\ }\href
  {https://doi.org/10.1088/0954-3899/19/1/002} {\bibfield  {journal} {\bibinfo
  {journal} {Journal of Physics G: Nuclear and Particle Physics}\ }\textbf
  {\bibinfo {volume} {19}},\ \bibinfo {pages} {1} (\bibinfo {year}
  {1993})}\BibitemShut {NoStop}%
\bibitem [{\citenamefont {Stoitsov}\ \emph {et~al.}(2013)\citenamefont
  {Stoitsov}, \citenamefont {Schunck}, \citenamefont {Kortelainen},
  \citenamefont {Michel}, \citenamefont {Nam}, \citenamefont {Olsen},
  \citenamefont {Sarich},\ and\ \citenamefont {Wild}}]{CPC184_Stoitsov}%
  \BibitemOpen
  \bibfield  {author} {\bibinfo {author} {\bibfnamefont {M.}~\bibnamefont
  {Stoitsov}}, \bibinfo {author} {\bibfnamefont {N.}~\bibnamefont {Schunck}},
  \bibinfo {author} {\bibfnamefont {M.}~\bibnamefont {Kortelainen}}, \bibinfo
  {author} {\bibfnamefont {N.}~\bibnamefont {Michel}}, \bibinfo {author}
  {\bibfnamefont {H.}~\bibnamefont {Nam}}, \bibinfo {author} {\bibfnamefont
  {E.}~\bibnamefont {Olsen}}, \bibinfo {author} {\bibfnamefont
  {J.}~\bibnamefont {Sarich}},\ and\ \bibinfo {author} {\bibfnamefont
  {S.}~\bibnamefont {Wild}},\ }\bibfield  {title} {\bibinfo {title} {Axially
  deformed solution of the {Skyrme-Hartree-Fock-Bogoliubov} equations using
  the transformed harmonic oscillator basis (II) {HFBTHO} v2.00d: A new version
  of the program},\ }\href
  {https://doi.org/https://doi.org/10.1016/j.cpc.2013.01.013} {\bibfield
  {journal} {\bibinfo  {journal} {Computer Physics Communications}\ }\textbf
  {\bibinfo {volume} {184}},\ \bibinfo {pages} {1592} (\bibinfo {year}
  {2013})}\BibitemShut {NoStop}%
\bibitem [{\citenamefont {Bartel}\ \emph {et~al.}(1982)\citenamefont {Bartel},
  \citenamefont {Quentin}, \citenamefont {Brack}, \citenamefont {Guet},\ and\
  \citenamefont {Håkansson}}]{NPA386_Bartel}%
  \BibitemOpen
  \bibfield  {author} {\bibinfo {author} {\bibfnamefont {J.}~\bibnamefont
  {Bartel}}, \bibinfo {author} {\bibfnamefont {P.}~\bibnamefont {Quentin}},
  \bibinfo {author} {\bibfnamefont {M.}~\bibnamefont {Brack}}, \bibinfo
  {author} {\bibfnamefont {C.}~\bibnamefont {Guet}},\ and\ \bibinfo {author}
  {\bibfnamefont {H.-B.}\ \bibnamefont {Håkansson}},\ }\bibfield  {title}
  {\bibinfo {title} {{Towards a better parametrisation of Skyrme-like effective
  forces: A critical study of the SkM force}},\ }\href
  {https://doi.org/https://doi.org/10.1016/0375-9474(82)90403-1} {\bibfield
  {journal} {\bibinfo  {journal} {Nuclear Physics A}\ }\textbf {\bibinfo
  {volume} {386}},\ \bibinfo {pages} {79} (\bibinfo {year} {1982})}\BibitemShut
  {NoStop}%
\bibitem [{\citenamefont {Schunck}\ \emph {et~al.}(2014)\citenamefont
  {Schunck}, \citenamefont {Duke}, \citenamefont {Carr},\ and\ \citenamefont
  {Knoll}}]{PRC90_Schunck}%
  \BibitemOpen
  \bibfield  {author} {\bibinfo {author} {\bibfnamefont {N.}~\bibnamefont
  {Schunck}}, \bibinfo {author} {\bibfnamefont {D.}~\bibnamefont {Duke}},
  \bibinfo {author} {\bibfnamefont {H.}~\bibnamefont {Carr}},\ and\ \bibinfo
  {author} {\bibfnamefont {A.}~\bibnamefont {Knoll}},\ }\bibfield  {title}
  {\bibinfo {title} {Description of induced nuclear fission with {Skyrme}
  energy functionals: {Static} potential energy surfaces and fission fragment
  properties},\ }\href {https://doi.org/10.1103/PhysRevC.90.054305} {\bibfield
  {journal} {\bibinfo  {journal} {Phys. Rev. C}\ }\textbf {\bibinfo {volume}
  {90}},\ \bibinfo {pages} {054305} (\bibinfo {year} {2014})}\BibitemShut
  {NoStop}%
\bibitem [{\citenamefont {Dobaczewski}\ \emph {et~al.}(2002)\citenamefont
  {Dobaczewski}, \citenamefont {Nazarewicz},\ and\ \citenamefont
  {Stoitsov}}]{EPJA15_Dobaczewski}%
  \BibitemOpen
  \bibfield  {author} {\bibinfo {author} {\bibfnamefont {J.}~\bibnamefont
  {Dobaczewski}}, \bibinfo {author} {\bibfnamefont {W.}~\bibnamefont
  {Nazarewicz}},\ and\ \bibinfo {author} {\bibfnamefont {M.~V.}\ \bibnamefont
  {Stoitsov}},\ }\bibfield  {title} {\bibinfo {title} {Nuclear ground-state
  properties from mean-field calculations},\ }\href
  {https://doi.org/10.1140/epja/i2001-10218-8} {\bibfield  {journal} {\bibinfo
  {journal} {The European Physical Journal A}\ }\textbf {\bibinfo {volume}
  {15}},\ \bibinfo {pages} {21} (\bibinfo {year} {2002})}\BibitemShut {NoStop}%
\bibitem [{\citenamefont {Abusara}\ \emph {et~al.}(2012)\citenamefont
  {Abusara}, \citenamefont {Afanasjev},\ and\ \citenamefont
  {Ring}}]{PRC85_Abusara}%
  \BibitemOpen
  \bibfield  {author} {\bibinfo {author} {\bibfnamefont {H.}~\bibnamefont
  {Abusara}}, \bibinfo {author} {\bibfnamefont {A.~V.}\ \bibnamefont
  {Afanasjev}},\ and\ \bibinfo {author} {\bibfnamefont {P.}~\bibnamefont
  {Ring}},\ }\bibfield  {title} {\bibinfo {title} {Fission barriers in
  covariant density functional theory: Extrapolation to superheavy nuclei},\
  }\href {https://doi.org/10.1103/PhysRevC.85.024314} {\bibfield  {journal}
  {\bibinfo  {journal} {Phys. Rev. C}\ }\textbf {\bibinfo {volume} {85}},\
  \bibinfo {pages} {024314} (\bibinfo {year} {2012})}\BibitemShut {NoStop}%
\bibitem [{\citenamefont {Scamps}\ \emph {et~al.}(2015)\citenamefont {Scamps},
  \citenamefont {Simenel},\ and\ \citenamefont {Lacroix}}]{PRC92_Scamps}%
  \BibitemOpen
  \bibfield  {author} {\bibinfo {author} {\bibfnamefont {G.}~\bibnamefont
  {Scamps}}, \bibinfo {author} {\bibfnamefont {C.}~\bibnamefont {Simenel}},\
  and\ \bibinfo {author} {\bibfnamefont {D.}~\bibnamefont {Lacroix}},\
  }\bibfield  {title} {\bibinfo {title} {Superfluid dynamics of
  $^{258}\mathrm{Fm}$ fission},\ }\href
  {https://doi.org/10.1103/PhysRevC.92.011602} {\bibfield  {journal} {\bibinfo
  {journal} {Phys. Rev. C}\ }\textbf {\bibinfo {volume} {92}},\ \bibinfo
  {pages} {011602} (\bibinfo {year} {2015})}\BibitemShut {NoStop}%
\bibitem [{\citenamefont {Zhao}\ \emph {et~al.}(2016)\citenamefont {Zhao},
  \citenamefont {Lu}, \citenamefont {Nik\ifmmode \check{s}\else
  \v{s}\fi{}i\ifmmode~\acute{c}\else \'{c}\fi{}}, \citenamefont {Vretenar},\
  and\ \citenamefont {Zhou}}]{PRC93_Zhao}%
  \BibitemOpen
  \bibfield  {author} {\bibinfo {author} {\bibfnamefont {J.}~\bibnamefont
  {Zhao}}, \bibinfo {author} {\bibfnamefont {B.-N.}\ \bibnamefont {Lu}},
  \bibinfo {author} {\bibfnamefont {T.}~\bibnamefont {Nik\ifmmode
  \check{s}\else \v{s}\fi{}i\ifmmode~\acute{c}\else \'{c}\fi{}}}, \bibinfo
  {author} {\bibfnamefont {D.}~\bibnamefont {Vretenar}},\ and\ \bibinfo
  {author} {\bibfnamefont {S.-G.}\ \bibnamefont {Zhou}},\ }\bibfield  {title}
  {\bibinfo {title} {Multidimensionally-constrained relativistic mean-field
  study of spontaneous fission: Coupling between shape and pairing degrees of
  freedom},\ }\href {https://doi.org/10.1103/PhysRevC.93.044315} {\bibfield
  {journal} {\bibinfo  {journal} {Phys. Rev. C}\ }\textbf {\bibinfo {volume}
  {93}},\ \bibinfo {pages} {044315} (\bibinfo {year} {2016})}\BibitemShut
  {NoStop}%
\bibitem [{\citenamefont {Warda}\ and\ \citenamefont
  {Egido}(2012)}]{PRC86_Warda}%
  \BibitemOpen
  \bibfield  {author} {\bibinfo {author} {\bibfnamefont {M.}~\bibnamefont
  {Warda}}\ and\ \bibinfo {author} {\bibfnamefont {J.~L.}\ \bibnamefont
  {Egido}},\ }\bibfield  {title} {\bibinfo {title} {Fission half-lives of
  superheavy nuclei in a microscopic approach},\ }\href
  {https://doi.org/10.1103/PhysRevC.86.014322} {\bibfield  {journal} {\bibinfo
  {journal} {Phys. Rev. C}\ }\textbf {\bibinfo {volume} {86}},\ \bibinfo
  {pages} {014322} (\bibinfo {year} {2012})}\BibitemShut {NoStop}%
\bibitem [{\citenamefont {Baran}\ \emph {et~al.}(2015)\citenamefont {Baran},
  \citenamefont {Kowal}, \citenamefont {Reinhard}, \citenamefont {Robledo},
  \citenamefont {Staszczak},\ and\ \citenamefont {Warda}}]{NPA944_Baran}%
  \BibitemOpen
  \bibfield  {author} {\bibinfo {author} {\bibfnamefont {A.}~\bibnamefont
  {Baran}}, \bibinfo {author} {\bibfnamefont {M.}~\bibnamefont {Kowal}},
  \bibinfo {author} {\bibfnamefont {P.-G.}\ \bibnamefont {Reinhard}}, \bibinfo
  {author} {\bibfnamefont {L.}~\bibnamefont {Robledo}}, \bibinfo {author}
  {\bibfnamefont {A.}~\bibnamefont {Staszczak}},\ and\ \bibinfo {author}
  {\bibfnamefont {M.}~\bibnamefont {Warda}},\ }\bibfield  {title} {\bibinfo
  {title} {Fission barriers and probabilities of spontaneous fission for
  elements with ${Z}\ge100$},\ }\href
  {https://doi.org/https://doi.org/10.1016/j.nuclphysa.2015.06.002} {\bibfield
  {journal} {\bibinfo  {journal} {Nuclear Physics A}\ }\textbf {\bibinfo
  {volume} {944}},\ \bibinfo {pages} {442} (\bibinfo {year}
  {2015})}\BibitemShut {NoStop}%
\bibitem [{\citenamefont {Ignatyuk}(1985)}]{Ignatyuk_book}%
  \BibitemOpen
  \bibfield  {author} {\bibinfo {author} {\bibfnamefont {A.}~\bibnamefont
  {Ignatyuk}},\ }\href {https://nds.iaea.org/publications/indc/indc-ccp-0233/}
  {\emph {\bibinfo {title} {Statistical Properties of Excited Atomic Nuclei}}}\
  (\bibinfo  {publisher} {International Atomic Energy Agency, Vienna},\ \bibinfo {year}
  {1985})\BibitemShut {NoStop}%
\bibitem [{\citenamefont {Rahmatinejad}\ \emph {et~al.}(2022)\citenamefont
  {Rahmatinejad}, \citenamefont {Shneidman}, \citenamefont {Adamian},
  \citenamefont {Antonenko}, \citenamefont {Jachimowicz},\ and\ \citenamefont
  {Kowal}}]{PRC105_Rahmatinejad}%
  \BibitemOpen
  \bibfield  {author} {\bibinfo {author} {\bibfnamefont {A.}~\bibnamefont
  {Rahmatinejad}}, \bibinfo {author} {\bibfnamefont {T.~M.}\ \bibnamefont
  {Shneidman}}, \bibinfo {author} {\bibfnamefont {G.~G.}\ \bibnamefont
  {Adamian}}, \bibinfo {author} {\bibfnamefont {N.~V.}\ \bibnamefont
  {Antonenko}}, \bibinfo {author} {\bibfnamefont {P.}~\bibnamefont
  {Jachimowicz}},\ and\ \bibinfo {author} {\bibfnamefont {M.}~\bibnamefont
  {Kowal}},\ }\bibfield  {title} {\bibinfo {title} {Energy dependent ratios of
  level-density parameters in superheavy nuclei},\ }\href
  {https://doi.org/10.1103/PhysRevC.105.044328} {\bibfield  {journal} {\bibinfo
   {journal} {Phys. Rev. C}\ }\textbf {\bibinfo {volume} {105}},\ \bibinfo
  {pages} {044328} (\bibinfo {year} {2022})}\BibitemShut {NoStop}%
\bibitem [{\citenamefont {Ignatyuk}\ \emph {et~al.}(1979)\citenamefont
  {Ignatyuk}, \citenamefont {Istekov},\ and\ \citenamefont
  {Smirenkin}}]{SovJNPhy29_Ignatyuk}%
  \BibitemOpen
  \bibfield  {author} {\bibinfo {author} {\bibfnamefont {A.~V.}\ \bibnamefont
  {Ignatyuk}}, \bibinfo {author} {\bibfnamefont {K.~K.}\ \bibnamefont
  {Istekov}},\ and\ \bibinfo {author} {\bibfnamefont {G.~N.}\ \bibnamefont
  {Smirenkin}},\ }\bibfield  {title} {\bibinfo {title} {Role of collective
  effects in systematics of level density of nuclei},\ }\href@noop {}
  {\bibfield  {journal} {\bibinfo  {journal} {Soviet Journal of Nuclear
  Physics}\ }\textbf {\bibinfo {volume} {29}},\ \bibinfo {pages} {450}
  (\bibinfo {year} {1979})}\BibitemShut {NoStop}%
\bibitem [{\citenamefont {Dobaczewski}\ \emph {et~al.}(2009)\citenamefont
  {Dobaczewski}, \citenamefont {Satuła}, \citenamefont {Carlsson},
  \citenamefont {Engel}, \citenamefont {Olbratowski}, \citenamefont
  {Powałowski}, \citenamefont {Sadziak}, \citenamefont {Sarich}, \citenamefont
  {Schunck}, \citenamefont {Staszczak}, \citenamefont {Stoitsov}, \citenamefont
  {Zalewski},\ and\ \citenamefont {Zduńczuk}}]{CPC_180_Dobaczewski}%
  \BibitemOpen
  \bibfield  {author} {\bibinfo {author} {\bibfnamefont {J.}~\bibnamefont
  {Dobaczewski}}, \bibinfo {author} {\bibfnamefont {W.}~\bibnamefont
  {Satuła}}, \bibinfo {author} {\bibfnamefont {B.}~\bibnamefont {Carlsson}},
  \bibinfo {author} {\bibfnamefont {J.}~\bibnamefont {Engel}}, \bibinfo
  {author} {\bibfnamefont {P.}~\bibnamefont {Olbratowski}}, \bibinfo {author}
  {\bibfnamefont {P.}~\bibnamefont {Powałowski}}, \bibinfo {author}
  {\bibfnamefont {M.}~\bibnamefont {Sadziak}}, \bibinfo {author} {\bibfnamefont
  {J.}~\bibnamefont {Sarich}}, \bibinfo {author} {\bibfnamefont
  {N.}~\bibnamefont {Schunck}}, \bibinfo {author} {\bibfnamefont
  {A.}~\bibnamefont {Staszczak}}, \bibinfo {author} {\bibfnamefont
  {M.}~\bibnamefont {Stoitsov}}, \bibinfo {author} {\bibfnamefont
  {M.}~\bibnamefont {Zalewski}},\ and\ \bibinfo {author} {\bibfnamefont
  {H.}~\bibnamefont {Zduńczuk}},\ }\bibfield  {title} {\bibinfo {title}
  {{Solution of the Skyrme-Hartree-Fock-Bogolyubov equations in the
  Cartesian deformed harmonic-oscillator basis.: (VI) HFODD (v2.40h): A new
  version of the program}},\ }\href
  {https://doi.org/https://doi.org/10.1016/j.cpc.2009.08.009} {\bibfield
  {journal} {\bibinfo  {journal} {Computer Physics Communications}\ }\textbf
  {\bibinfo {volume} {180}},\ \bibinfo {pages} {2361} (\bibinfo {year}
  {2009})}\BibitemShut {NoStop}%
\bibitem [{\citenamefont {Schunck}\ \emph {et~al.}(2012)\citenamefont
  {Schunck}, \citenamefont {Dobaczewski}, \citenamefont {McDonnell},
  \citenamefont {Satuła}, \citenamefont {Sheikh}, \citenamefont {Staszczak},
  \citenamefont {Stoitsov},\ and\ \citenamefont {Toivanen}}]{CPC183_Schunck}%
  \BibitemOpen
  \bibfield  {author} {\bibinfo {author} {\bibfnamefont {N.}~\bibnamefont
  {Schunck}}, \bibinfo {author} {\bibfnamefont {J.}~\bibnamefont
  {Dobaczewski}}, \bibinfo {author} {\bibfnamefont {J.}~\bibnamefont
  {McDonnell}}, \bibinfo {author} {\bibfnamefont {W.}~\bibnamefont {Satuła}},
  \bibinfo {author} {\bibfnamefont {J.}~\bibnamefont {Sheikh}}, \bibinfo
  {author} {\bibfnamefont {A.}~\bibnamefont {Staszczak}}, \bibinfo {author}
  {\bibfnamefont {M.}~\bibnamefont {Stoitsov}},\ and\ \bibinfo {author}
  {\bibfnamefont {P.}~\bibnamefont {Toivanen}},\ }\bibfield  {title} {\bibinfo
  {title} {{Solution of the Skyrme-Hartree-Fock-Bogolyubov equations in
  the Cartesian deformed harmonic-oscillator basis.: (VII) HVODD (v2.49t): A
  new version of the program}},\ }\href
  {https://doi.org/https://doi.org/10.1016/j.cpc.2011.08.013} {\bibfield
  {journal} {\bibinfo  {journal} {Computer Physics Communications}\ }\textbf
  {\bibinfo {volume} {183}},\ \bibinfo {pages} {166} (\bibinfo {year}
  {2012})}\BibitemShut {NoStop}%
\end{thebibliography}

%

\end{document}